\documentclass[]{spie}  
\usepackage[]{graphicx}
\usepackage[]{hyperref}
\def\rd{$^{\rm rd}$}
\def\th{$^{\rm th}$}

\title{Three recipes for improving the image quality with optical long-baseline interferometers: BFMC, LFF, \& DPSC} 

\author{Florentin A. Millour\supit{a} and Martin Vannier\supit{a} and Anthony Meilland\supit{a}
\skiplinehalf
\supit{a} Laboratoire Lagrange, UMR7293, Universit\'e de Nice Sophia-Antipolis, CNRS, Observatoire de la C\^ote dÕAzur, Bd. de lÕObservatoire, 06304 Nice, France \\
}

\authorinfo{Further author information: (Send correspondence to F. Millour)\\F. Millour: E-mail: fmillour@oca.eu, Telephone: +33 (0)4 92 00 30 68\\  
}

\begin{document} 
  \maketitle 

\begin{abstract}
We present here three recipes for getting better images with optical interferometers. Two of them, Low-Frequencies Filling and Brute-Force Monte Carlo were used in our participation to the Interferometry Beauty Contest this year and can be applied to classical imaging using $V^2$ and closure phases. These two addition to image reconstruction provide a way of having more reliable images. The last recipe is similar in its principle as the self-calibration technique used in radio-interferometry. We call it also Ç self-calibration È, but it uses the wavelength-differential phase as a proxy of the object phase to build-up a full-featured complex visibility set of the observed object. This technique needs a first image-reconstruction run with an available software, using closure-phases and squared visibilities only. We used it for two scientific papers with great success. We discuss here the pros and cons of such imaging technique.
\end{abstract}


\keywords{Optical long-baseline interferometry, image reconstruction, inverse problem}

\section{Introduction}
\label{sec:intro}  


Imaging with optical interferometry is a somewhat new application, giving access to images of stellar surfaces or inner stellar environments.
Image reconstruction is a non-convex inverse problem, very tricky and complicated to cope with\cite{Thiebaut2010a}. The main reason of this situation is related to the sparsity of the spatial frequencies sampling by the interferometer\cite{Renard2011}. In the previous years, several algorithms have been developed, aiming at solving that ill-posed inverse problem of image-reconstruction, among which BBM\cite{2009A&A...497..195K, Millour2009a}, BSMEM\cite{2008SPIE.7013E.121B, Millour2009a}, MACIM\cite{Ireland2006b}, MIRA\cite{Millour2009a, Renard2010a, Benisty2011}, or WISARD\cite{LeBesnerais2008, Haubois2009}. These software do manage to produce decent images of astrophysical objects, but they are still giving a somewhat questionable image fidelity, mainly due to the small number of available telescopes simultaneously.

To cope with these strong limitations, we propose here three home-made recipes for getting better images. These recipes are relatively efficient regarding image artifacts, which are the main quality default seen on reconstructed image. These recipes are namely:
\begin{description}
\item[BFMC:] Brute-Force Monte Carlo method, described in section~\ref{secBFMC}
\item[LFF:] Low-Frequencies Filling, described in section~\ref{secLFF}
\item[DPSC:] Differential Phase Self-Calibration, described in section~\ref{secDPSC}.
\end{description}
The two first recipes presented here were used in our MIRA participation to the Interferometry Beauty Contest this year\cite{Baron2012}. The last recipe was published in Millour et al. 2011\cite{2011A&A...526A.107M}, where it improved significantly the image reconstruction quality of a circumstellar disk around a super-giant star.

\section{Brute-Force Monte Carlo (BFMC)}
\label{secBFMC}

Being located close to Monaco, we started to investigate the Monte Carlo method by adding random processes into existing image reconstruction software, to see if that could improve somewhat the final image quality. This work started with the MIRA software, the one where we have the most experience up to now. In principle, such methods could also be applied to BSMEM or WISARD.

While Monte Carlo methods, like Markov Chains, are used within the MACIM software for optimization, MIRA, BSMEM or WISARD do not use them by default. Instead, the classical way is to use Newton or quasi-Newton optimization. One way of introducing Monte Carlo into the imaging process is to act on the initial image. Therefore, we propose to input Brute-Force Monte Carlo (hereafter BFMC) into MIRA by randomly setting the initial image (a random noise plus either nothing, a Gaussian, or a uniform disk, with random size), and let MIRA reconstruct an image. Repeating this process a hundred times (a thousand times if the user has enough time) will scan the parameters space of the image reconstruction process, and provide a set of converged images with different solutions. An analysis is done afterwards on the results, in order to generate the final image.

This post-processing analysis is based on the convergence of the final image to the data by computing a $\chi^2$. Images are sorted by increasing $\chi^2$, a centering is made by computing the photo-center around the brightest pixel of the image, and a median image is computed on the best $\chi^2$ images. We also compute a map of standard deviation (RMS) between the different reconstructed images, which provide a proxy for error estimates of the final map. 

To illustrate the effect of the presented recipes, we came back to older data, used in Millour et al. 2009\cite{Millour2009a}, that were used to detect a companion star to the supergiant B[e] star HD87643. Supergiant B[e] stars exhibit the "B[e] phenomenon", i.e. a combination of emission lines from Hydrogen and metallic elements, both allowed and forbidden, plus a large infrared excess associated with dust in the vicinity of the star. Finding a companion star to HD87643 shed a new light to this complex system. In the 2009 paper, however, we could not constrain the shape of the extended circumbinary material, and we investigate here if the new methods we are developing could help us in constraining it.

\begin{figure}[htbp]
\centering
\begin{tabular}{ccc}
MIRA alone&MIRA + BFMC&RMS map\\
\includegraphics[totalheight=0.3\textwidth, angle=-90]{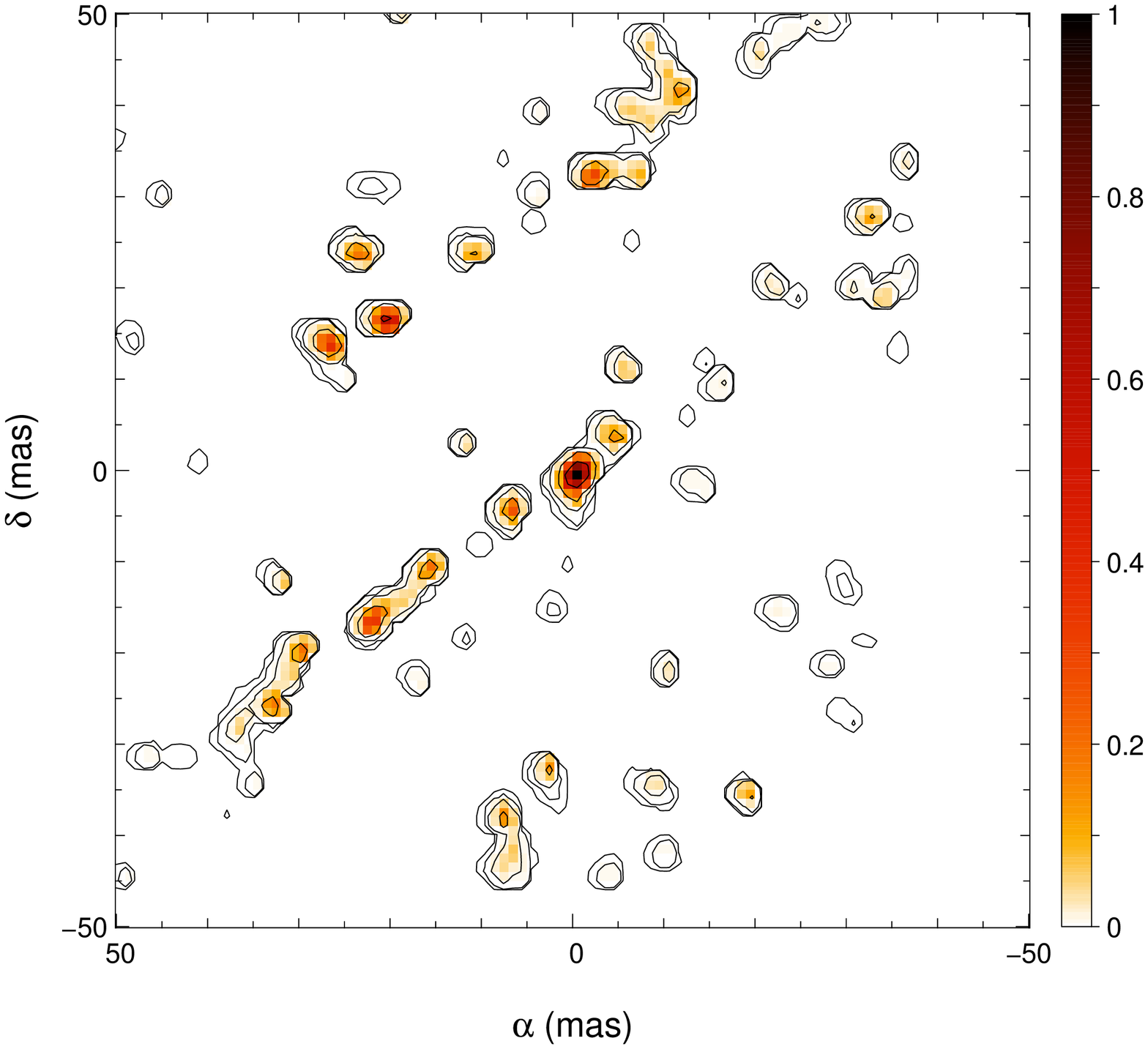}&
\includegraphics[totalheight=0.3\textwidth, angle=-90]{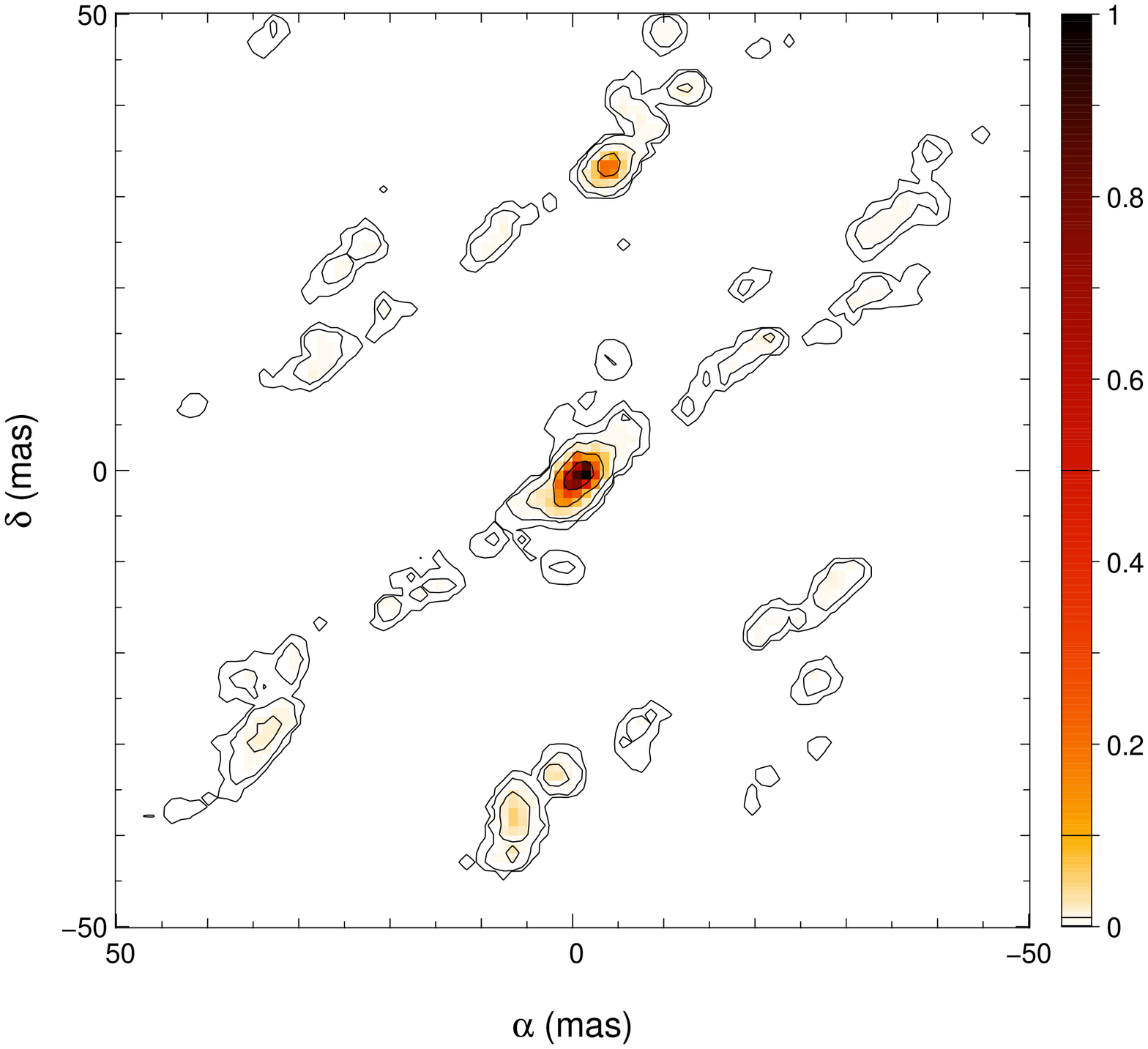}&
\includegraphics[totalheight=0.3\textwidth, angle=-90]{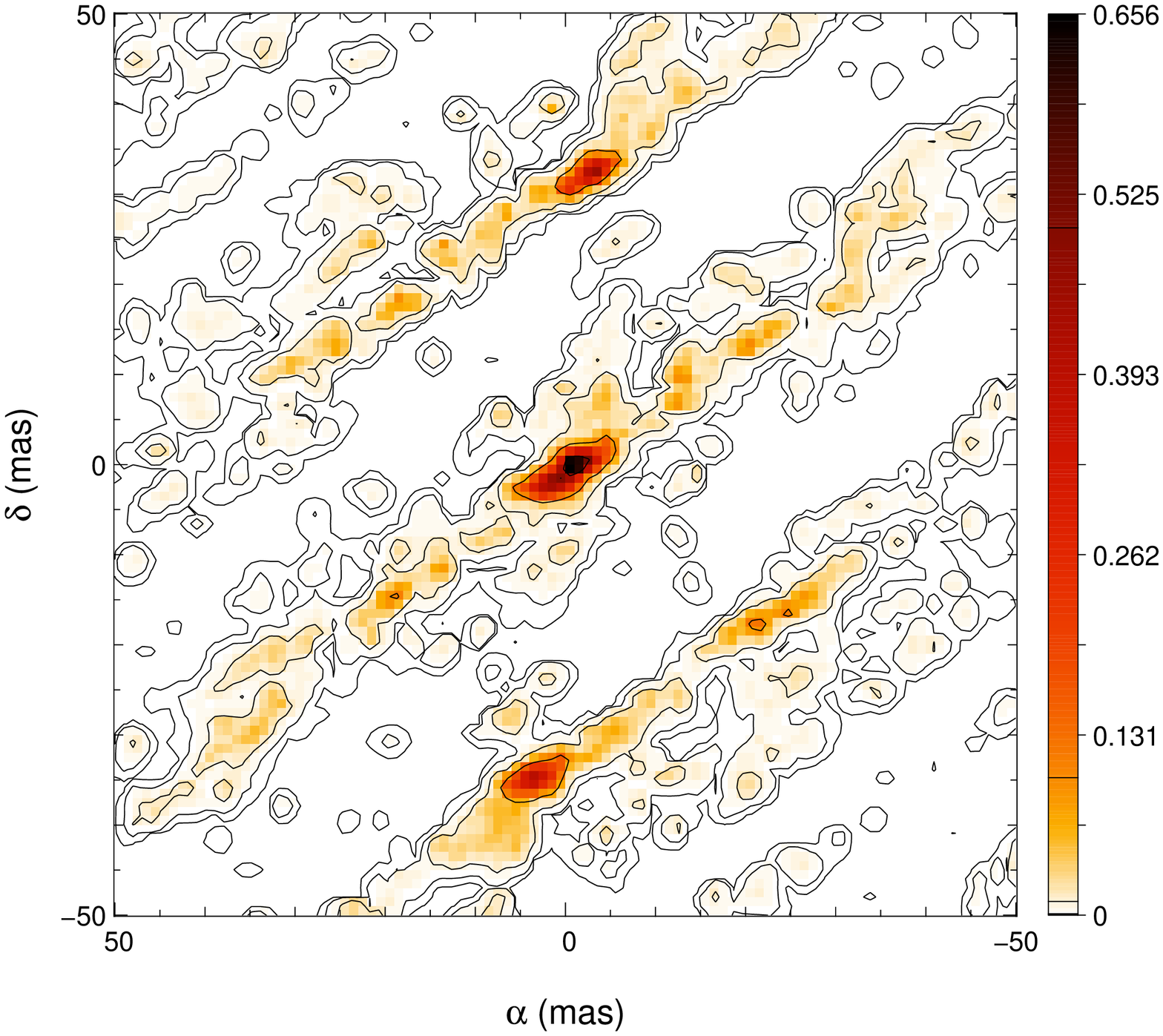}
\end{tabular}
\smallskip
\caption{Left: Example image reconstruction run based on the data of Millour et al. 2009\cite{Millour2009a}, made using the MIRA software and a random initial image (that image can be compared to the Figure 2 of Millour et al. 2009). Right: After BFMC with 100 random initial images, and taking the median image of the selected best $\chi^2$. All images have contours at 50\%, 10\%, 1\%, 0.1\% and 0.01\% of the peak intensity, to emphasize the artifacts.}
\label{imgHD87643}
\end{figure}

We reconstructed 100 images with random initial images constructed as a random noise plus either nothing, a Gaussian, or a uniform disk, with random size. We applied the post-processing described above.
The results can be seen in Figure~\ref{imgHD87643}, where we used MIRA with a maximum entropy regularization. The BFMC-obtained image (middle panel) can be compared to a MIRA imaging run with classical parameters (on the left).

Compared to the classical way of reconstructing images (left part of Figure~\ref{imgHD87643}), the presented BFMC method (middle panel) decreases significantly the number and amplitude of image artifacts (all the spurious dots seen in the left image). However, it does not wash out the main artifacts (here: oblique stripes in the image), and we also note that it slightly decreases the image spatial resolution, due to the fact that the centering process has a somewhat variable precision from individual image to image. We also note that the final image looks somewhat symmetric. This is also due to the centering, which sometimes focus on the wrong source, producing the apparently repeated companion star in the South of the system. This might be improved in the future using a smarter centroiding algorithm, or by validating (and possibly correcting) this step by hand.

An interesting output of this method is the RMS map, which is computed on the final set of images (see right panel of Figure~\ref{imgHD87643}). This map emphasizes the artifact stripes, which are due to the incomplete UV coverage in the N-E direction (see the corresponding UV coverage in Figure~\ref{V2HD87643SpFreq}).

\section{Low-Frequencies Filling (LFF)}
\label{secLFF}

We are faced today, with current imaging software, to final images which contain lots of artifacts, due to the UV coverage holes of the interferometer (see for example Figure~\ref{imgHD87643}), even when using the method presented in Section~\ref{secBFMC}.
Previous attempts to cope with this issue, in addition to formal regularizations, were using hard limitations of the field of view of the image by setting to zero all pixels outside a given zone \cite{Monnier2007a}, or reconstructing centro-symmetric objects when symmetries are known to exist in the image of the object \cite{LeBouquin2009c, Chiavassa2010}. Our proposition here is to go beyond these two ideas by applying what we call "Low Frequencies Filling" (hereafter LFF).

The principle of LFF is inspired from the work of radio-interferometrists\cite{2004A&A...419L..35B}, where it has been demonstrated that getting the low spatial frequencies via large collecting antennae, in addition to long-baseline interferometric data, provide a huge improvement on the image quality and image fidelity compared to standard imaging. In optical long-baseline interferometry, we usually do not have such low-frequency information available, except in a few cases\cite{Millour2009c}. Our proposition here is to fill these low-frequencies by using the available information at slightly higher frequency. To do so, we need \emph{at least} a few data points in the first visibility lobe, on which we can fit a model. The recipe is made in the three following steps:

\begin{description}
\item[Model-fitting on low frequencies: ] We select the low-spatial frequency part of the dataset to limit the analysis to the first visibility lobe, and we fit a model of the object to this subset of data. The model used can be composed of an increasing number of components (Gaussian, uniform disks, etc.) if the $\chi^2$ value of the fit is too high.
\item[Simulation of synthetic data: ] We perform a simulation, as if we observed the model fitted previously through an interferometer, with the following parameters:
\begin{itemize}
\item The number of baselines is the same as in the observation, but the largest baseline is set so as its spatial frequency does not go beyond the first visibility lobe,
\item The baselines are set randomly by fixing one telescope and randomly setting the positions of the other telescopes, in order to keep the same baseline organization as in the original file set,
\item The data is simulated using the previously-obtained model, and error bars are set in accordance with the error bars present in the original dataset. These steps allow feeding this simulated dataset into imaging software sensitive to the data format, like e.g. WISARD.
\end{itemize}
\item[Image reconstruction: ] The image reconstruction takes place using standard software (we tested this method with MIRA, BSMEM and WISARD), but instead of providing it the original dataset, we provide the original + synthetic dataset altogether.
\end{description}

\begin{figure}[htbp]
\centering
\begin{tabular}{cccc}
\includegraphics[height=0.18\textwidth, angle=-0, origin=br]{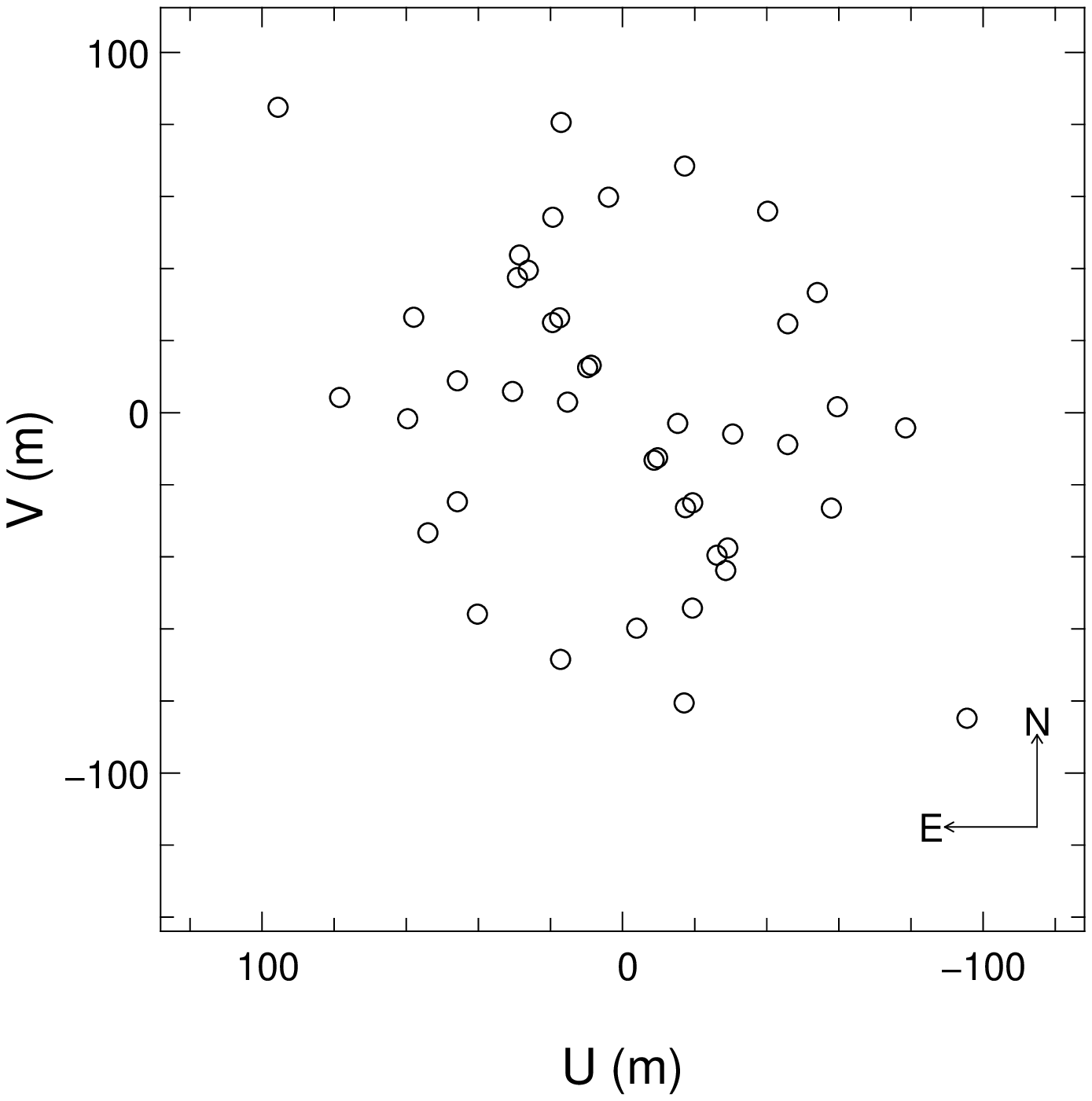}&
\includegraphics[height=0.28\textwidth, angle=-90, origin=br]{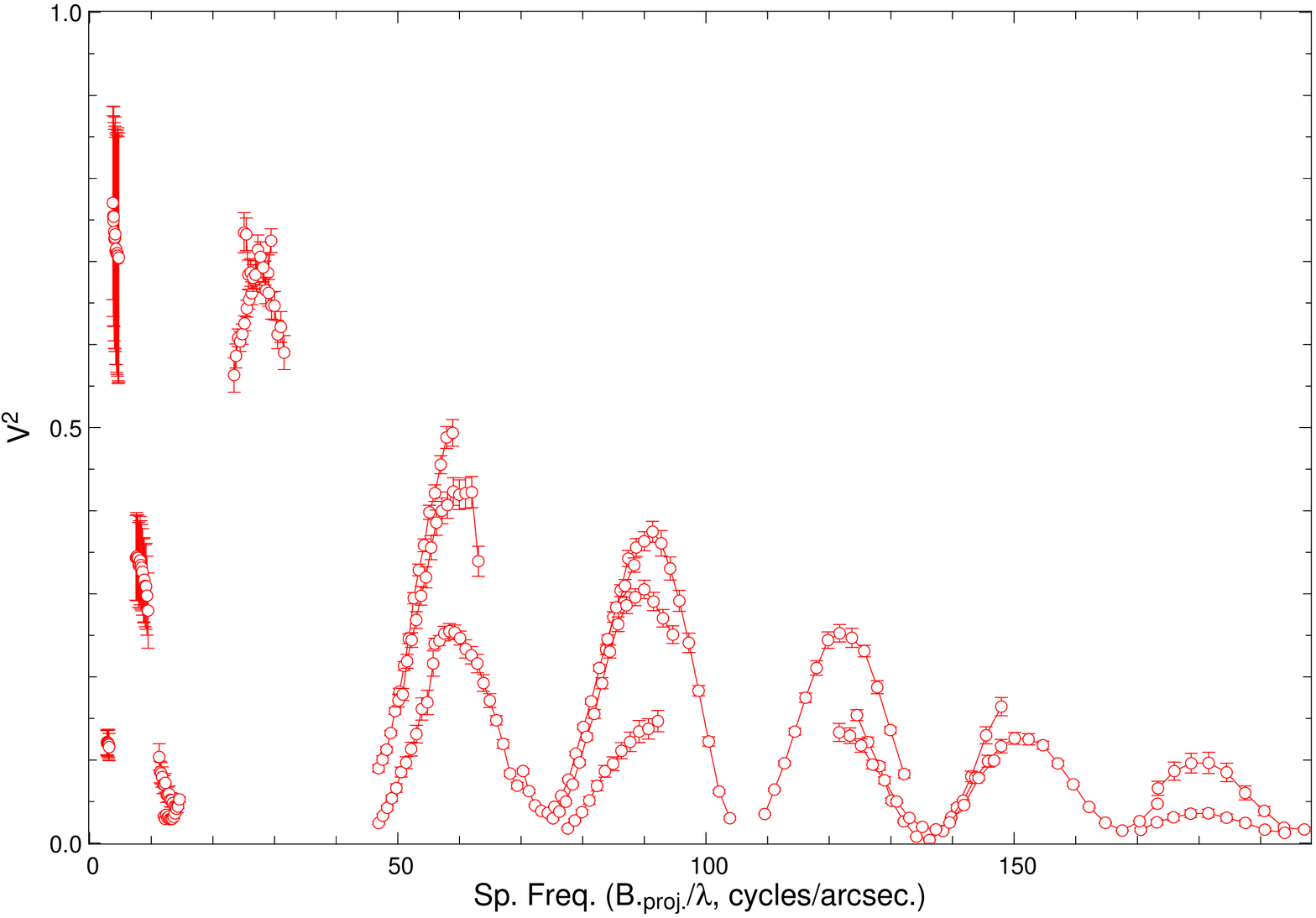}&
\includegraphics[height=0.18\textwidth, angle=-0, origin=br]{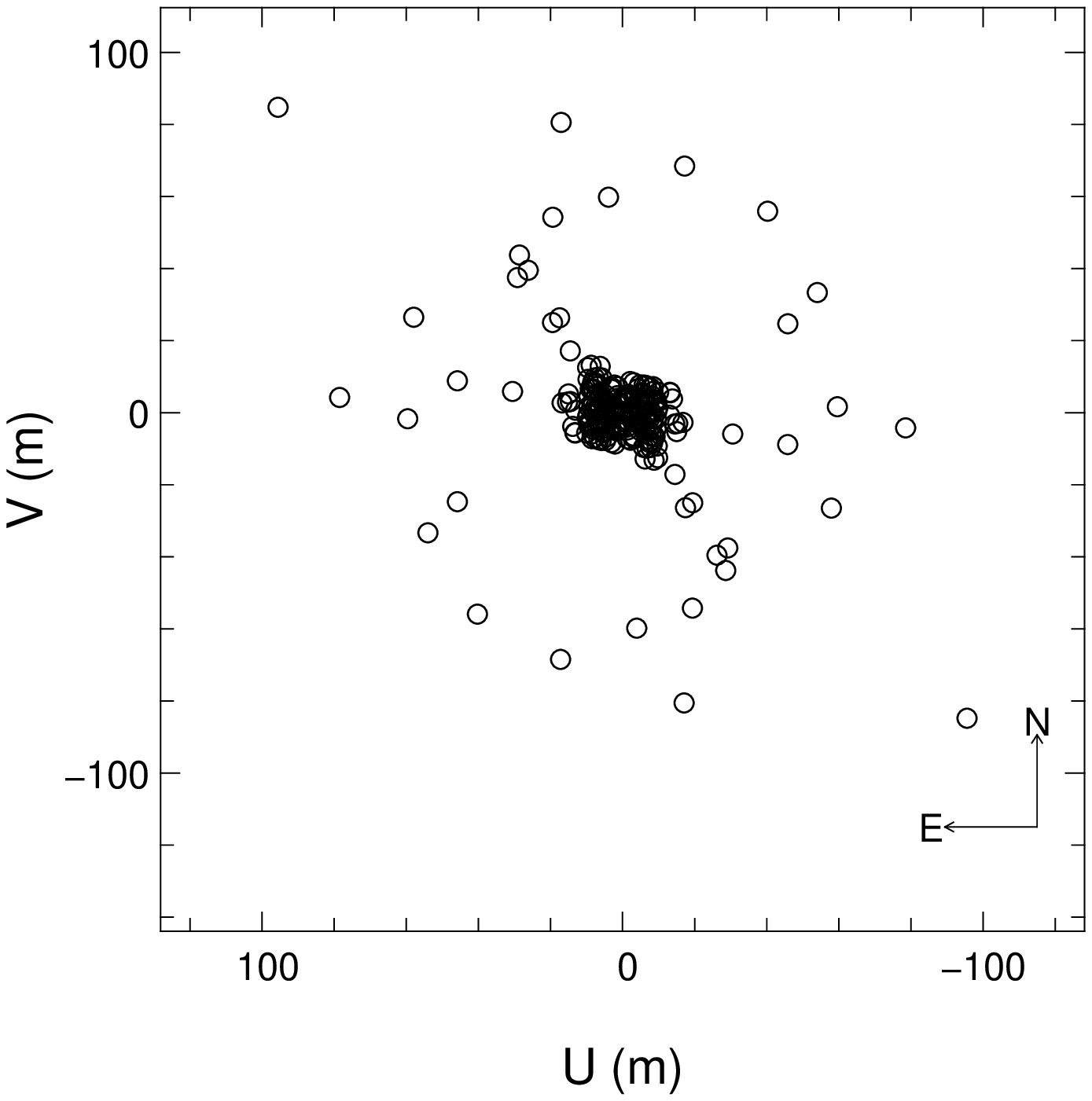}&
\includegraphics[height=0.28\textwidth, angle=-90, origin=br]{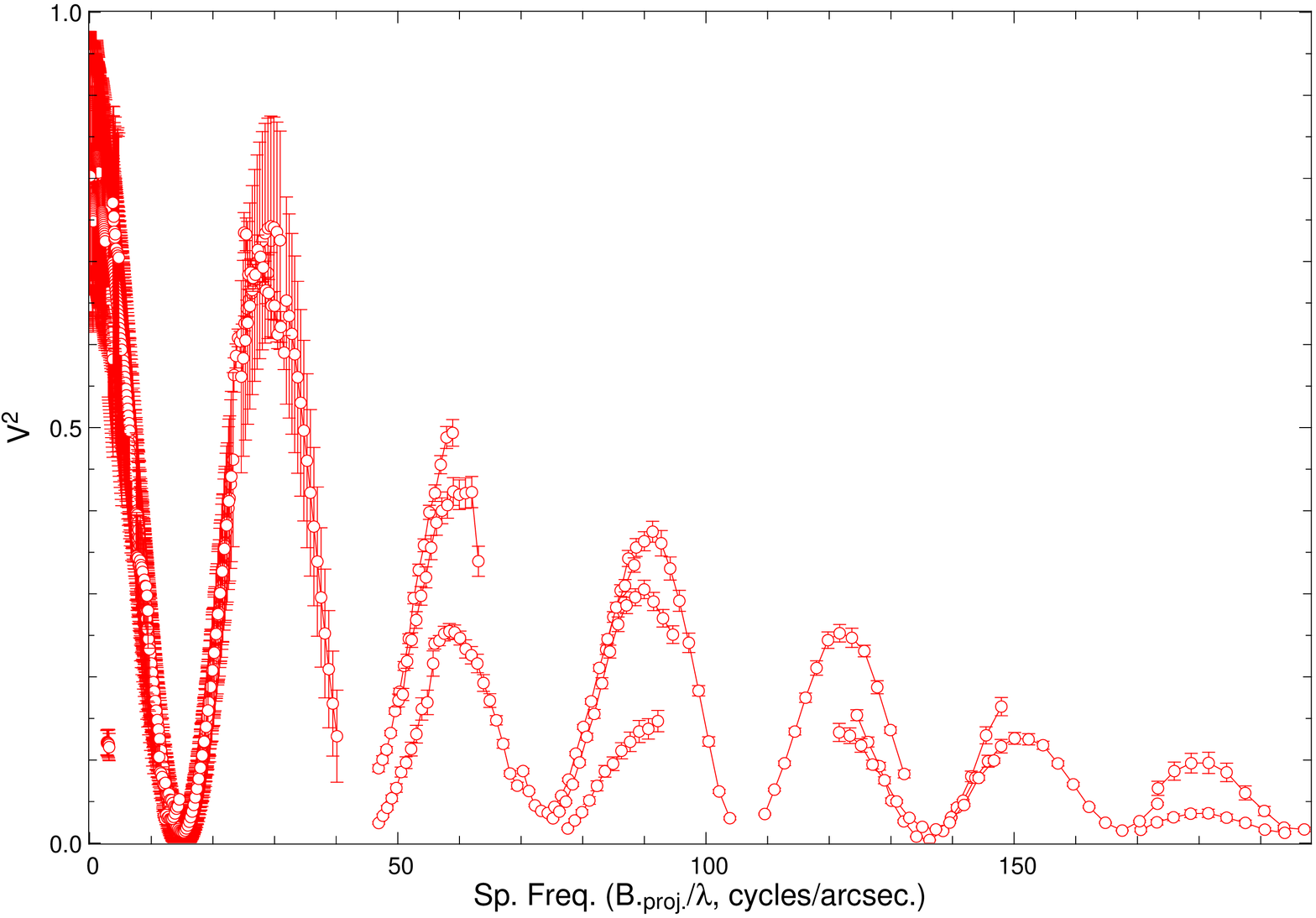}
\end{tabular}
\smallskip
\caption{Left: K-band squared visibilities of HD87643, as published in Millour et al. (2009)\cite{Millour2009a} together with the UV coverage (the visibilities are projected onto the main orientation of the binary star in order to exhibit the sine modulation of $V^2$). Right: Same but with the additional LFF synthetic data (note the additional points at low spatial frequencies).} 
\label{V2HD87643SpFreq}
\end{figure}

The great advantage of this idea compared to the others presented before is that we do not need to make strong assumptions on the space brightness distribution of the object (flux inside a zone, no flux outside), except from finding a simple geometrical model which fits the low-frequency data.
An example of the simulation process is illustrated in Figure~\ref{V2HD87643SpFreq}, where we added synthetic visibilities to the HD87643 dataset used in the previous section.

\begin{figure}[htbp]
\centering
\begin{tabular}{ccc}
MIRA +LFF & MIRA + LFF + BFMC&RMS map\\
\includegraphics[totalheight=0.3\textwidth, angle=-90, origin=br]{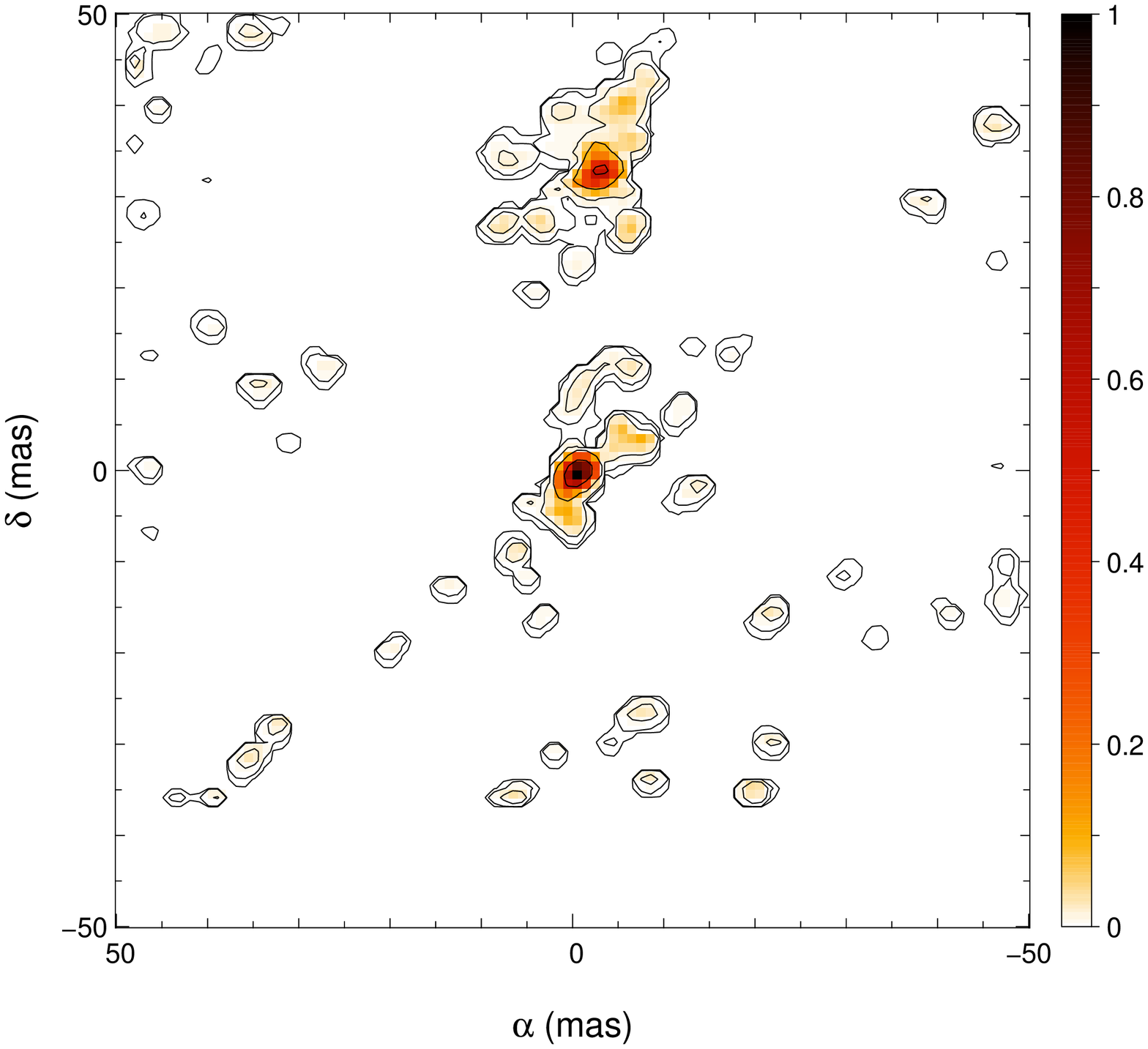}&
\includegraphics[totalheight=0.3\textwidth, angle=-90, origin=br]{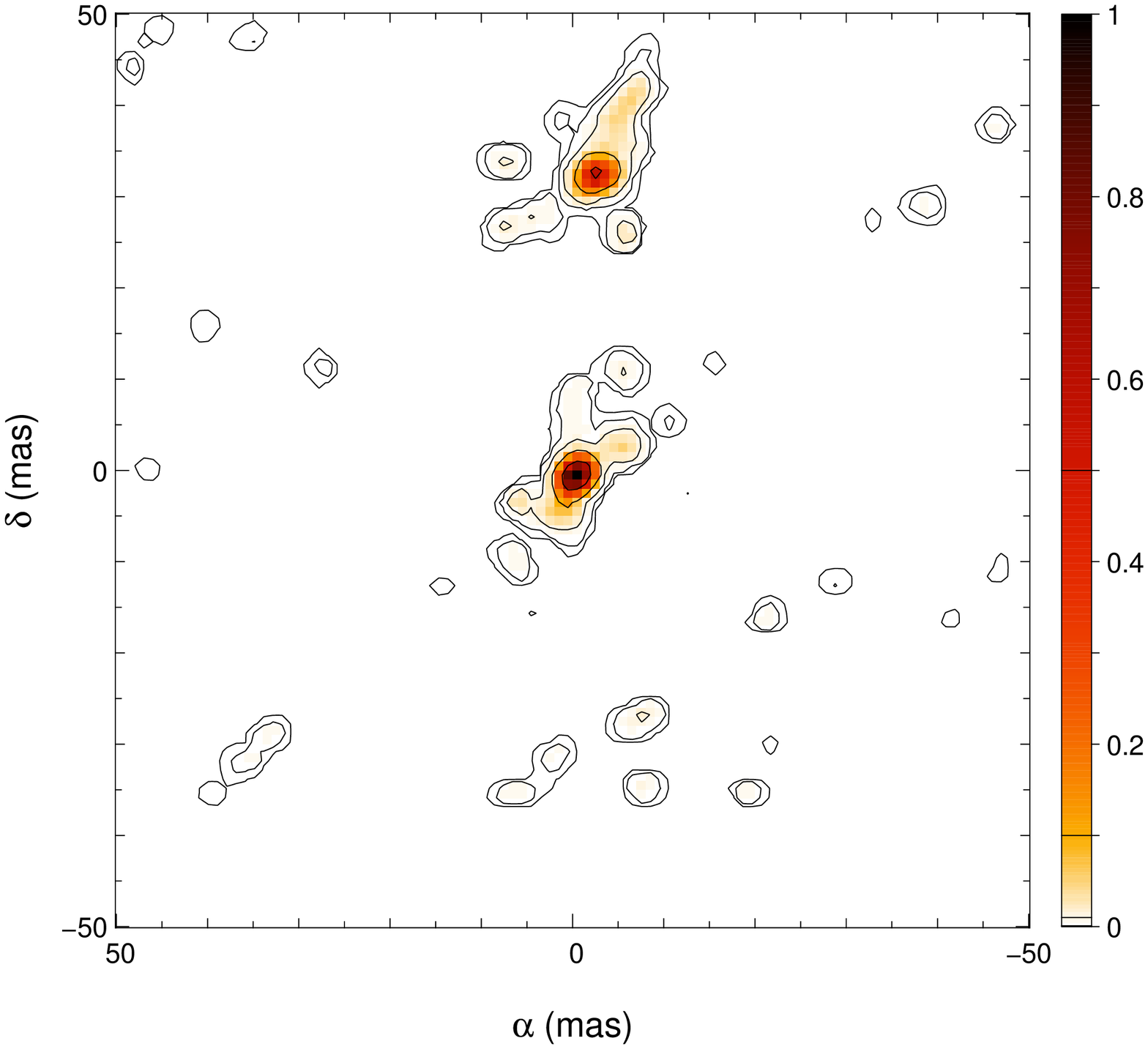}&
\includegraphics[totalheight=0.3\textwidth, angle=-90,origin=br]{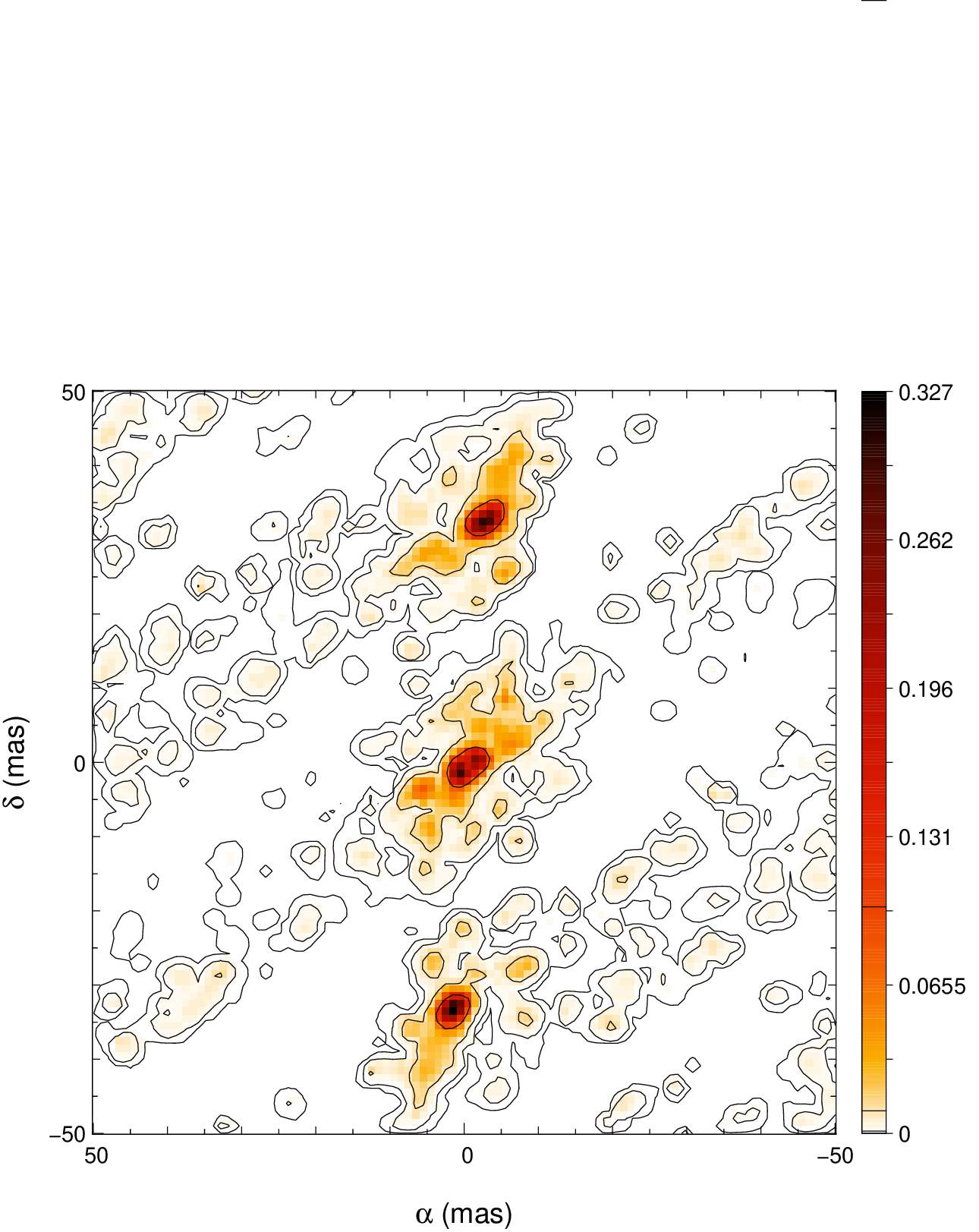}
\end{tabular}
\smallskip
\caption{Left: Same as Figure~\ref{imgHD87643}, this time using LFF.}
\label{imgHD87643LFF}
\end{figure}

The result of the image reconstruction with MIRA, with and without BFMC, is shown in Figure~\ref{imgHD87643LFF}.
We see a clear improvement of the images compared to Figure~\ref{imgHD87643}, including compared to the reconstructed image using BFMC alone. The "stripes" present in Figure~\ref{imgHD87643} have almost disappeared, and the RMS map has a much lower peak value, indicating the image reconstruction process is much more stable than in the previous section. However, in the case of HD87643, this does not help very much on locating the extended emission of the circumbinary material. We would need real low-frequency data (with the use of NACO/SAM for example), to definitively conclude on this material.

The combination of BFMC and LFF improves significantly the image reconstruction quality, except in a few cases (the Alp Fak object of the beauty contest this year is an example, where the centering process involved in the BFMC failed). These two recipes are easy to use on fast and scriptable image reconstruction software like MIRA. We are happily willing to provide the scripts used to implement these two recipes on demand.

\section{Differential phase Self-Calibration (DPSC)}
\label{secDPSC}

The last image reconstruction recipe we present here is somewhat different, because it needs the so-called wavelength-differential phase.

Differential phase was first used in long-baseline interferometry on the \emph{Grand Interf\'erom\`etre \`a deux T\'elescopes} (GI2T), with some success \cite{1995A&A...300..219S, 1997A&A...323..183V, 1998A&A...335..261V, 1999A&A...345..203B}. A cross-correlation technique was used to compute a phase as a function of wavelength on the GI2T.
Today, differential phase has been popularized by the Very Large Telescope Interferometer (VLTI), with both AMBER \cite{Petrov2007} and MIDI \cite{1998AGM....14..L05G} instruments, joined by the Navy Precision Optical Interferometer (NPOI, the "prototype" aspect having disappeared now) in its differential phase referencing mode \cite{2009ApJ...691..984S}, VEGA/CHARA \cite{2009A&A...508.1073M}, and Keck used in self-referenced phase mode \cite{2012PASP..124...51W}.

We describe in this section the differential phase and why it is interesting to input it in an imaging process, the proposition we made in Millour et al 2011\cite{2011A&A...526A.107M}, and a few examples showing the behavior of the Differential Phase Self-Calibration (hereafter DPSC).

\subsection{The differential phase} 

The main idea behind computing interferometry observable is to get the \emph{invariant} part of the wandering fringes of an interferometer, as seen through the atmosphere and a spectrograph. 
For example, one can get the Fourier amplitudes using speckle techniques (i.e. using the power spectrum instead of directly the Fourier transform of the fringes). This calculation provides $V^2$ (squared FT amplitude) as a function of wavelength.

However, for computing the Fourier phases, things get more complicated:
\begin{itemize}
\item One way is to use the closure relation when using three telescopes, the very same way as for bispectrum speckle interferometry \cite{1993A&A...278..328H}. One phase (out of three) can therefore be extracted, independent from any instrumental or atmospheric effect. This phase is called "closure phase", but when using 3 telescopes, it misses $2/3$ of the available phase information.
\item Another way is to have at his disposal a model of the \emph{variable} part of these wandering fringes, that can be subtracted from the raw Fourier Transforms before averaging. One does not need to have a model of the \emph{observed object}, but rather a model of the \emph{atmosphere} and the \emph{instrument} to correct for these effects and calculate a phase (called "differential phase"). A by-product of this method is that one can get also the Fourier amplitude (called "linear visibility", or "differential visibility").
\end{itemize}

This last point is what is called \emph{wavelength-differential} interferometry (often shortened \emph{differential} interferometry), because the instrument + atmosphere model is wavelength-dependent, and the observable are computed through a difference between the data and this model. The used model is time \emph{and} wavelength-dependent, and can be of growing complexity, depending on the wavelength coverage and spectral resolution of the instrument. The direct consequence is that this differential phase contains only a fraction of the original phase information.

This was illustrated in a few papers\cite{Millour2006d, Millour2008b, 2009ApJ...691..984S}, where a Taylor-Expansion of the observed phase $\phi_{ij}^{m}(\sigma)$ as a function of wave-number $\sigma = {^1}/{_\lambda}$ was presented to illustrate the information kept into the differential phase, and the one discarded by the process. Indeed, the observed phase, in the absence of nanometer-accuracy metrology, is affected by an unknown OPD term $\delta_{ij}(t)$, and higher order terms, which can come for example from the water-vapor and dry air chromatic dispersion $s_{ij}(t)$ :

\begin{equation}
  \phi^{m}_{ij}(t,\sigma) = \phi^{*}_{ij} (\sigma) + 2 \pi
    \delta_{ij}(t)  \sigma + 2 \pi
   s_{ij}(t)  \sigma^2 + ... + o(\sigma^n)
\label{eq:obsPhase}
\end{equation}

where $o(\sigma^n)$ is a negligible (or time-constant) term compared to all the other ones.
On the other side, one can perform a Taylor-Expansion of the object phase at the observation time:

\begin{equation}
  \phi_{ij}^{*}(\sigma)=a^{*}_{0}+a^{*}_{1}\sigma+a^{*}_{1}\sigma^2  + ... +
  \delta\varphi_{ij}^{*}(\sigma)
\label{eq:dvptPhase}
\end{equation}

where $ \delta\varphi_{ij}^{*}(\sigma)$ is the differential phase.
One can see here the similarities in the terms in in Eq.~\ref{eq:obsPhase} and \ref{eq:dvptPhase}. The differential data analysis process make disappear the 1-to-$n^{\rm th}$ first terms of this Taylor-Expansion. $n$ depends on the wavelength coverage and spectral resolution of the instrument. The different data processing processes are stacking the different contributions that can be modeled, depending on the spectral resolution. The higher the spectral resolution, the smaller will be $n$, because the smaller the variations of $\sigma$, making more and more of the higher-order terms negligible in the process.

\subsection{The self-calibration} 

The basic principle of self-calibration applied to spectro-interferometry is largely inspired from the work of radio-interferometrists\cite{1984ARA&A..22...97P}. We presented an application of this method in Millour et al. 2011\cite{2011A&A...526A.107M}, where we found it was improving the image reconstruction by a large factor, providing the missing astrometric information compared to reconstructing images with $V^2$ and closure phases only. Here, we illustrate the performances of this technique on a simulated dataset based on a synthetic Be star model.

\begin{figure}[htbp]
\centering
\begin{tabular}{ccccc}
Continuum & Blue wing & Center & Red wing\\
\multicolumn{4}{c}{Model}\\
\includegraphics[height=0.19\textwidth, angle=-0, origin=br]{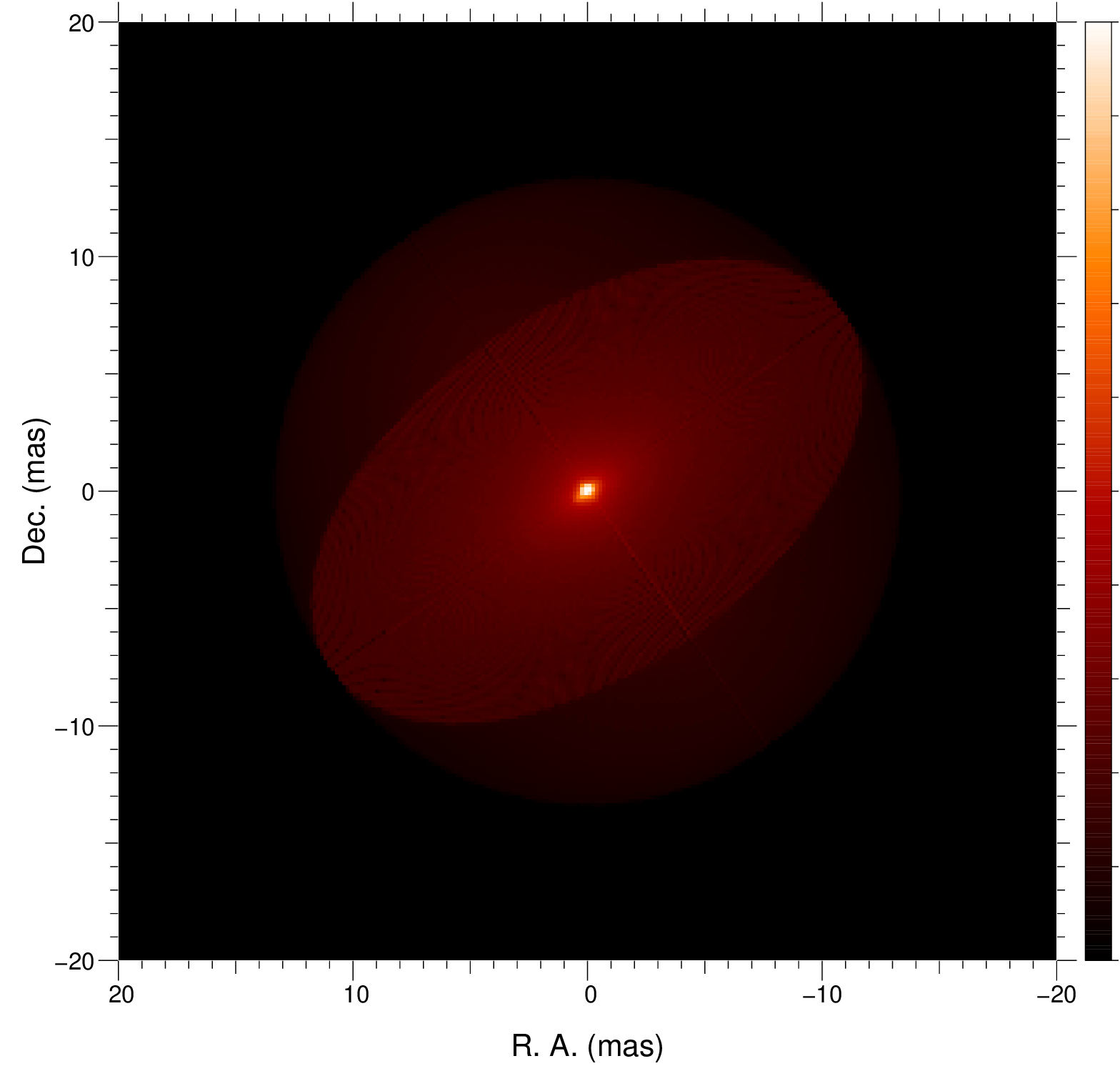}&
\includegraphics[height=0.19\textwidth, angle=-0, origin=br]{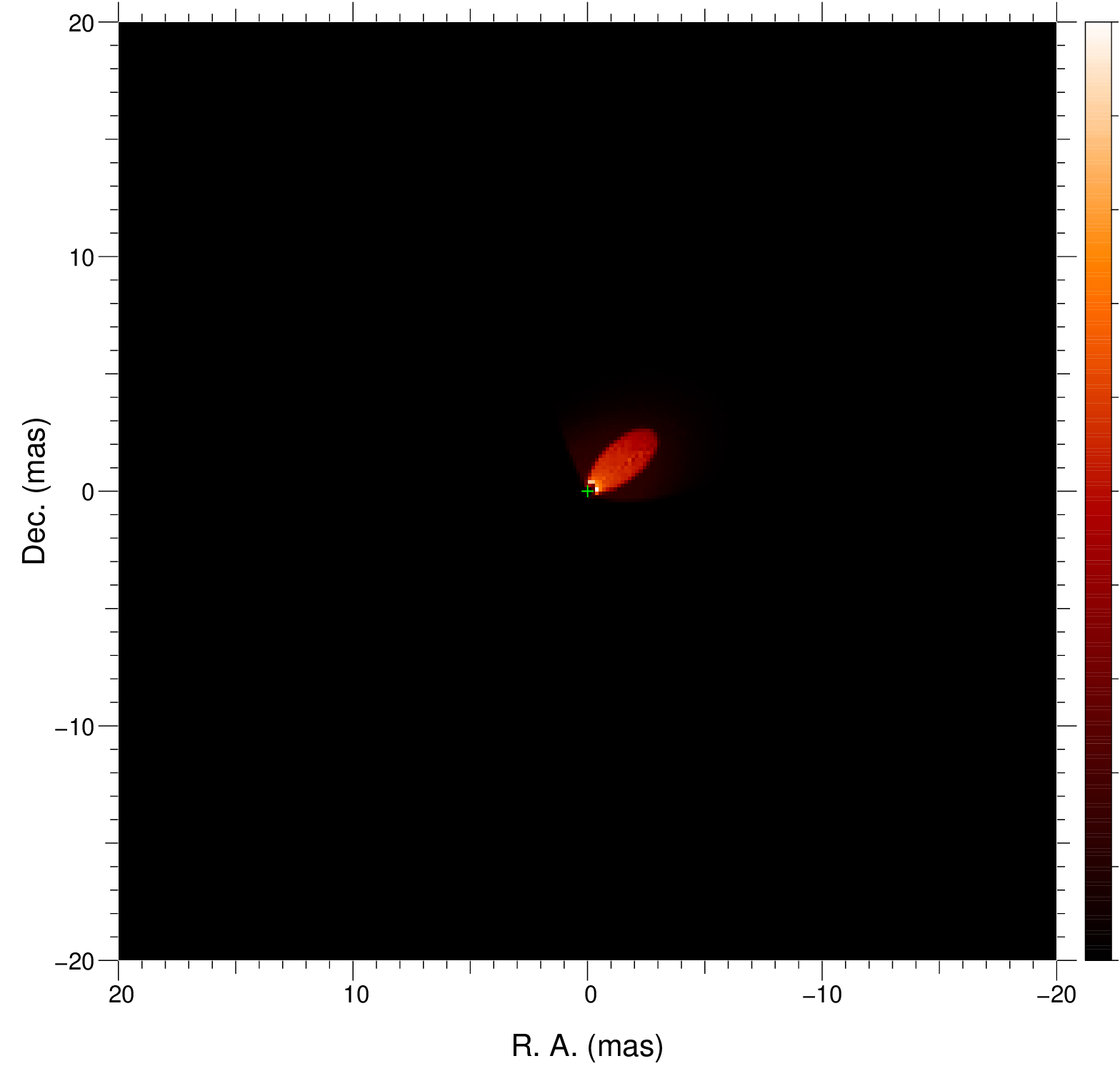}&
\includegraphics[height=0.19\textwidth, angle=-0,origin=br]{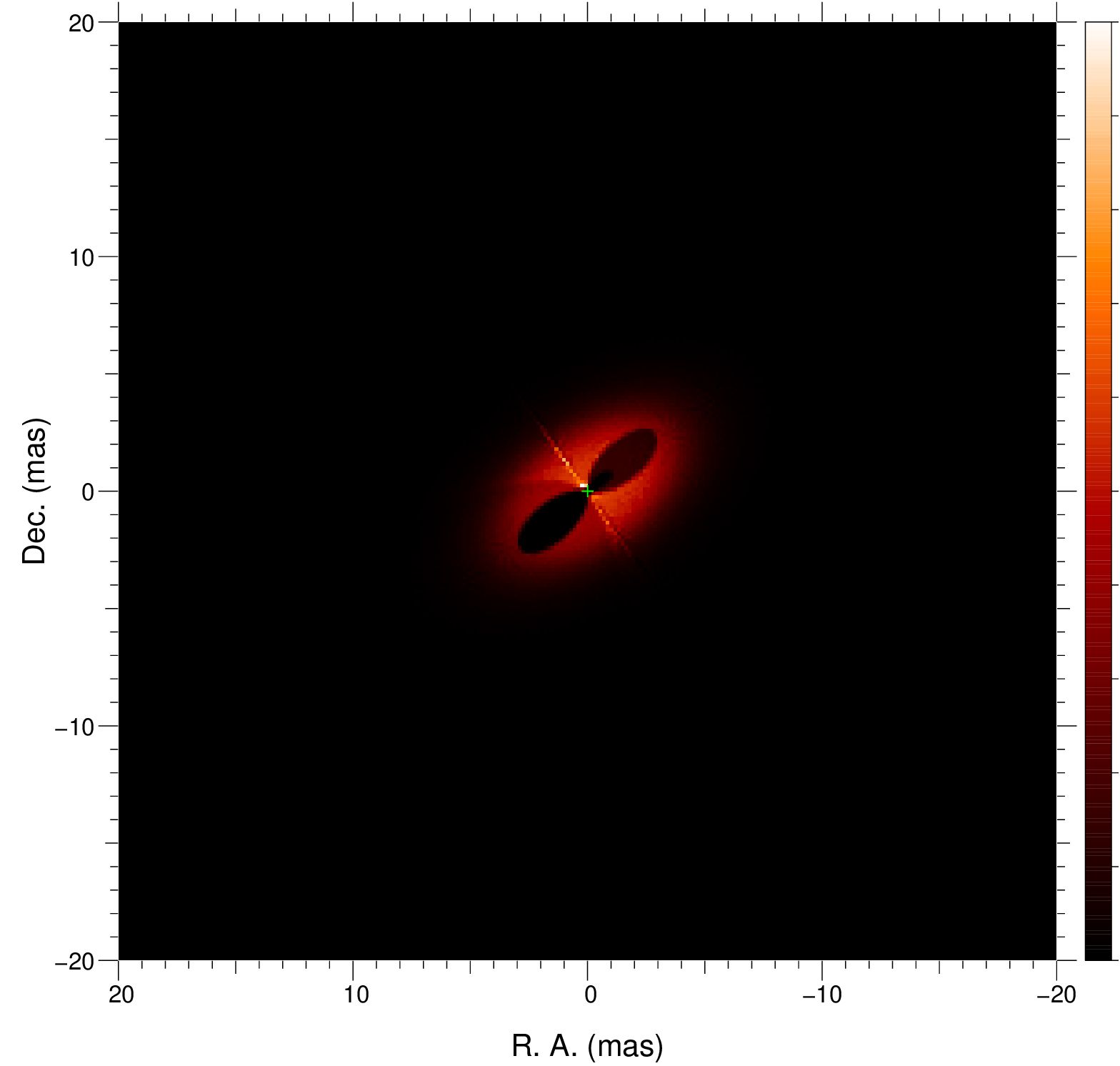}&
\includegraphics[height=0.19\textwidth, angle=-0,origin=br]{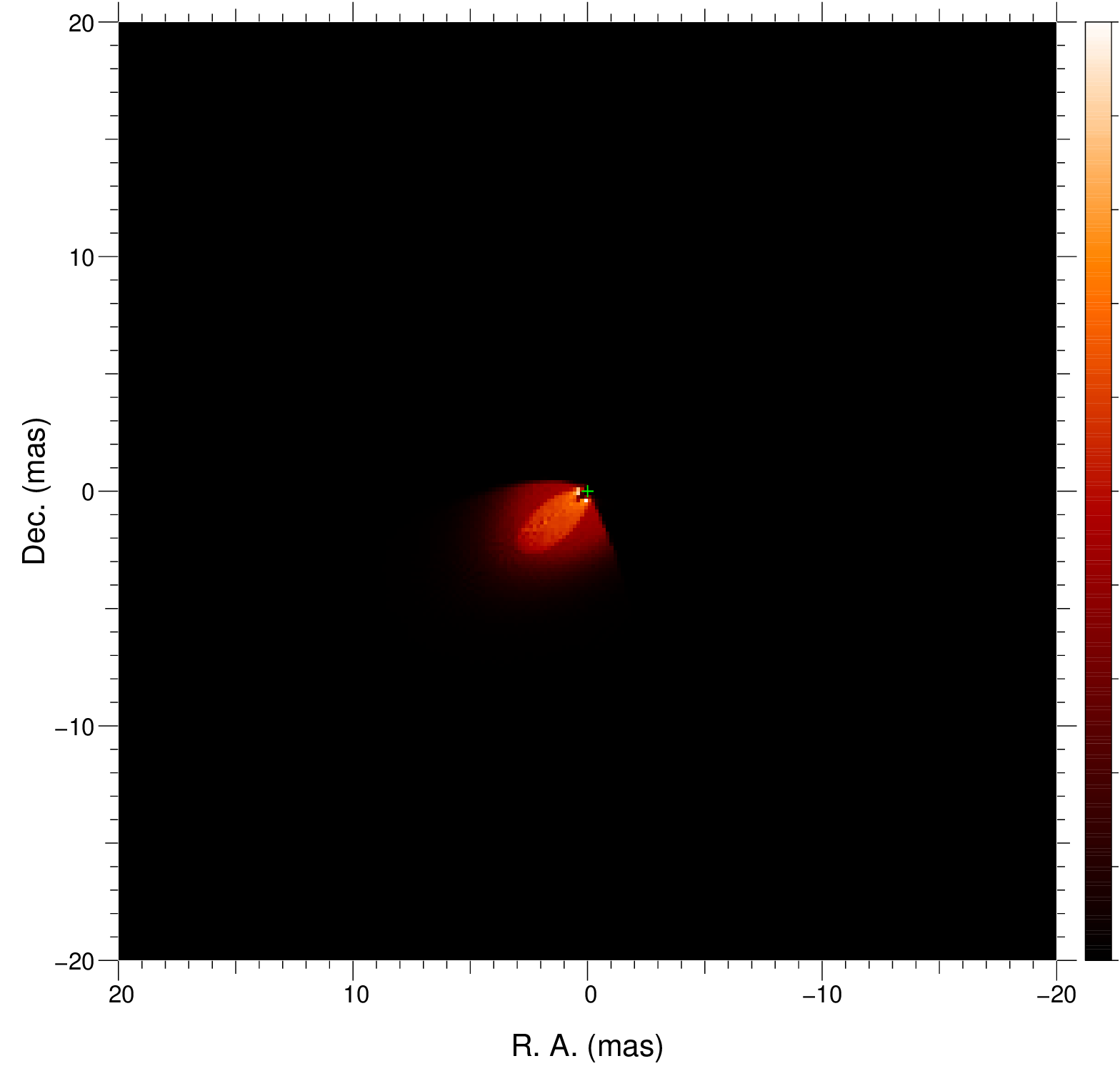}\\
\multicolumn{4}{c}{Image reconstruction ($V^2$ + closure phase)}\\
\includegraphics[height=0.19\textwidth, angle=-90, origin=br]{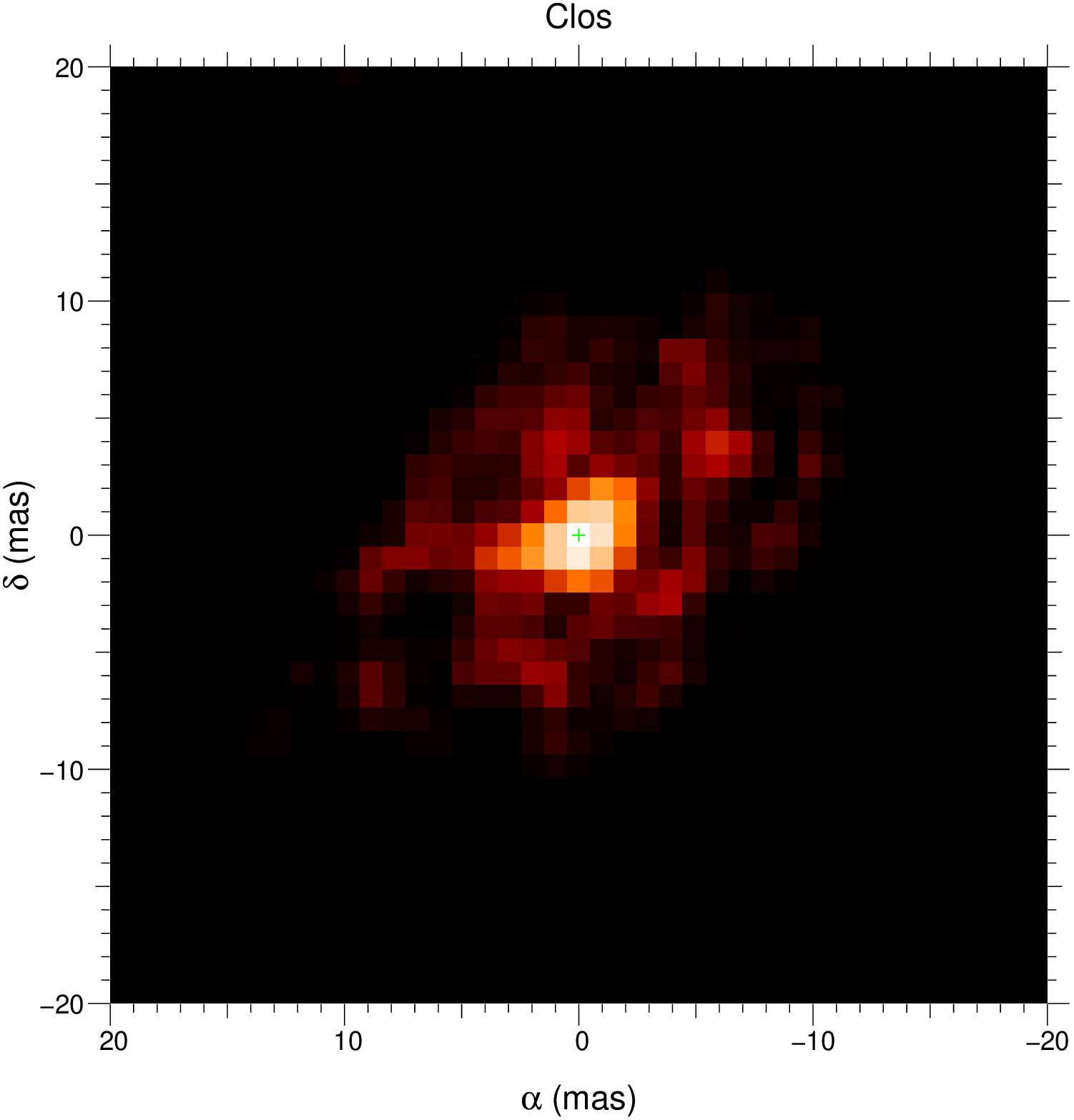}&
\includegraphics[height=0.19\textwidth, angle=-90, origin=br]{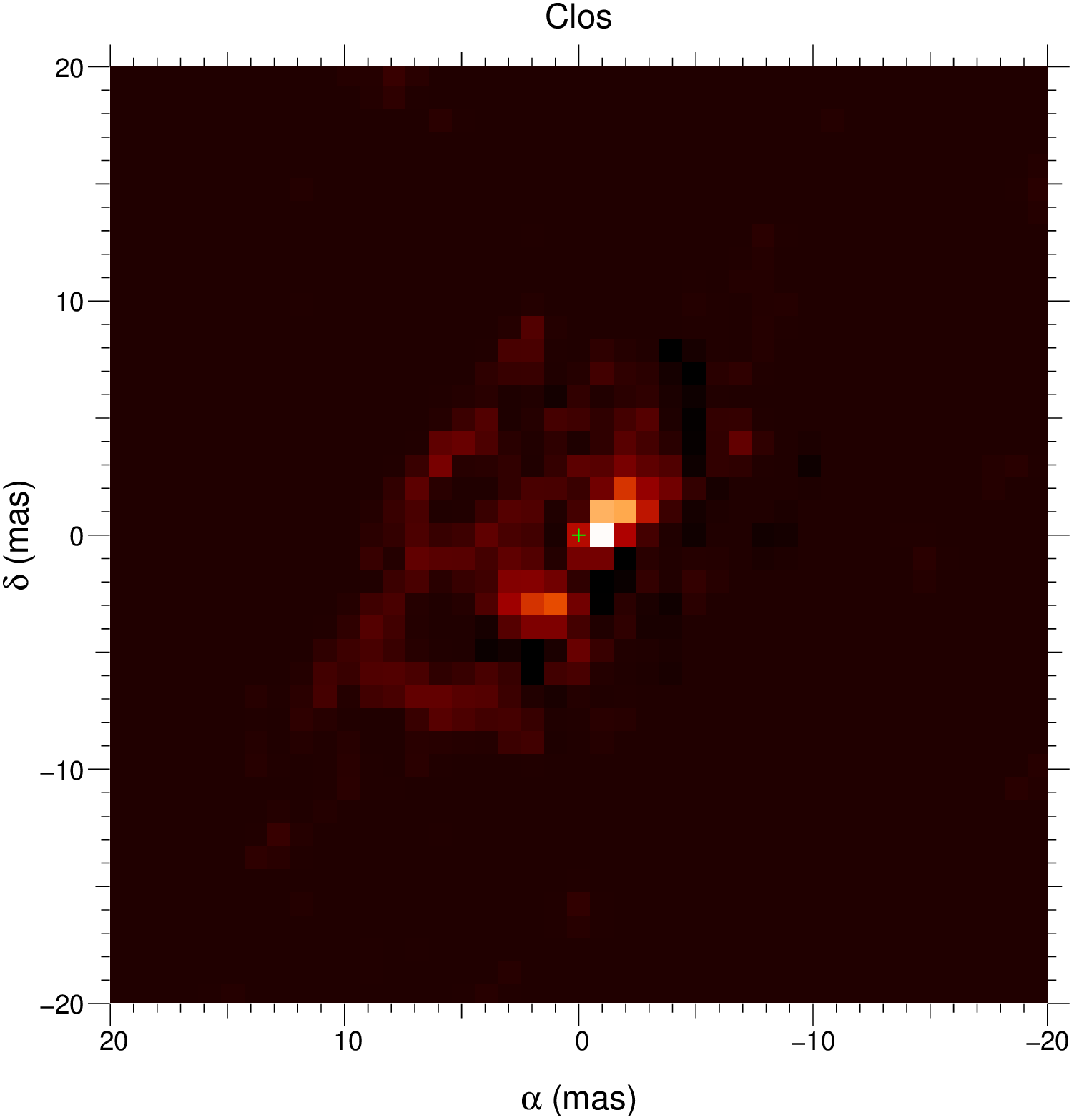}&
\includegraphics[height=0.19\textwidth, angle=-90,origin=br]{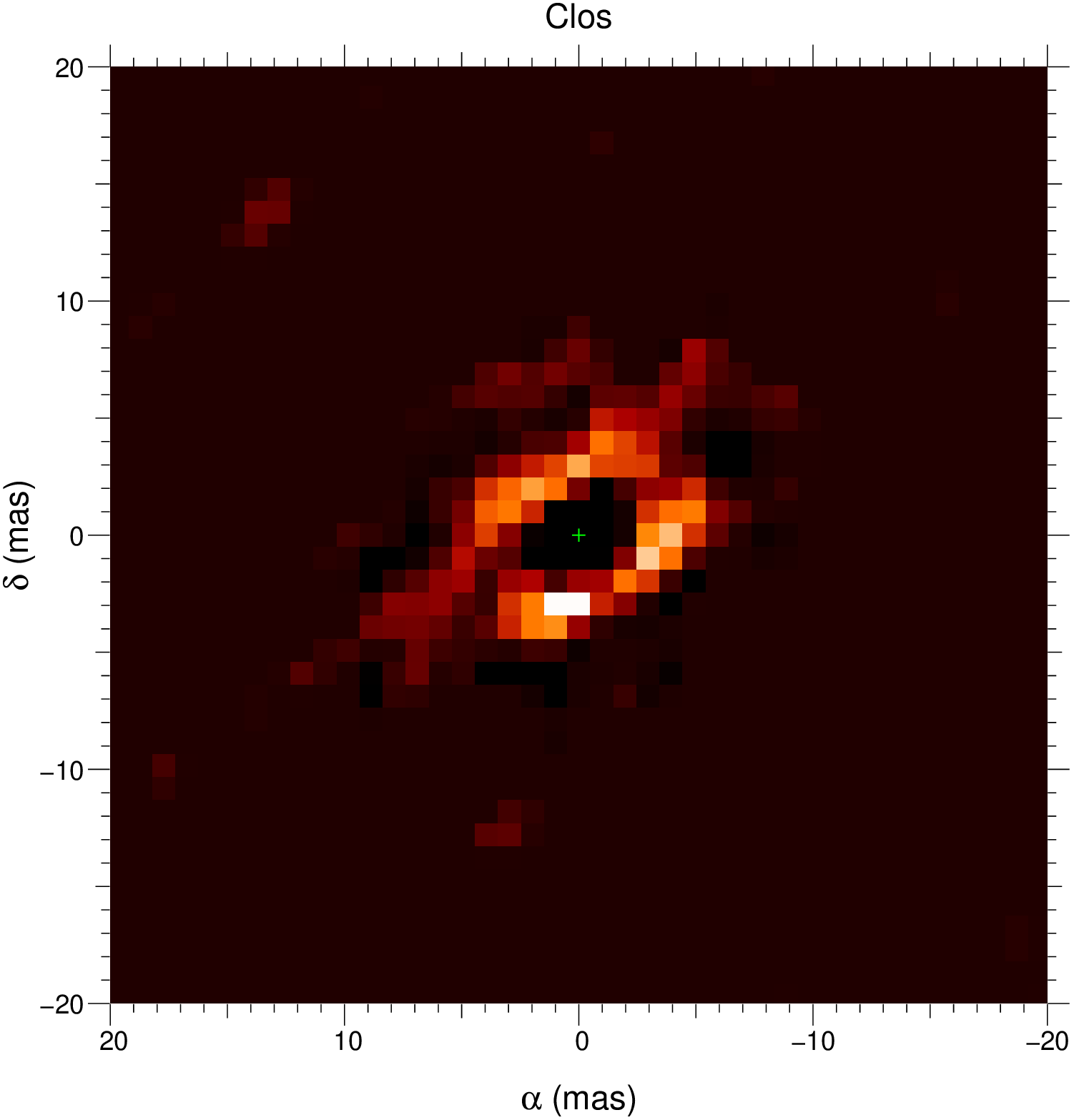}&
\includegraphics[height=0.19\textwidth, angle=-90,origin=br]{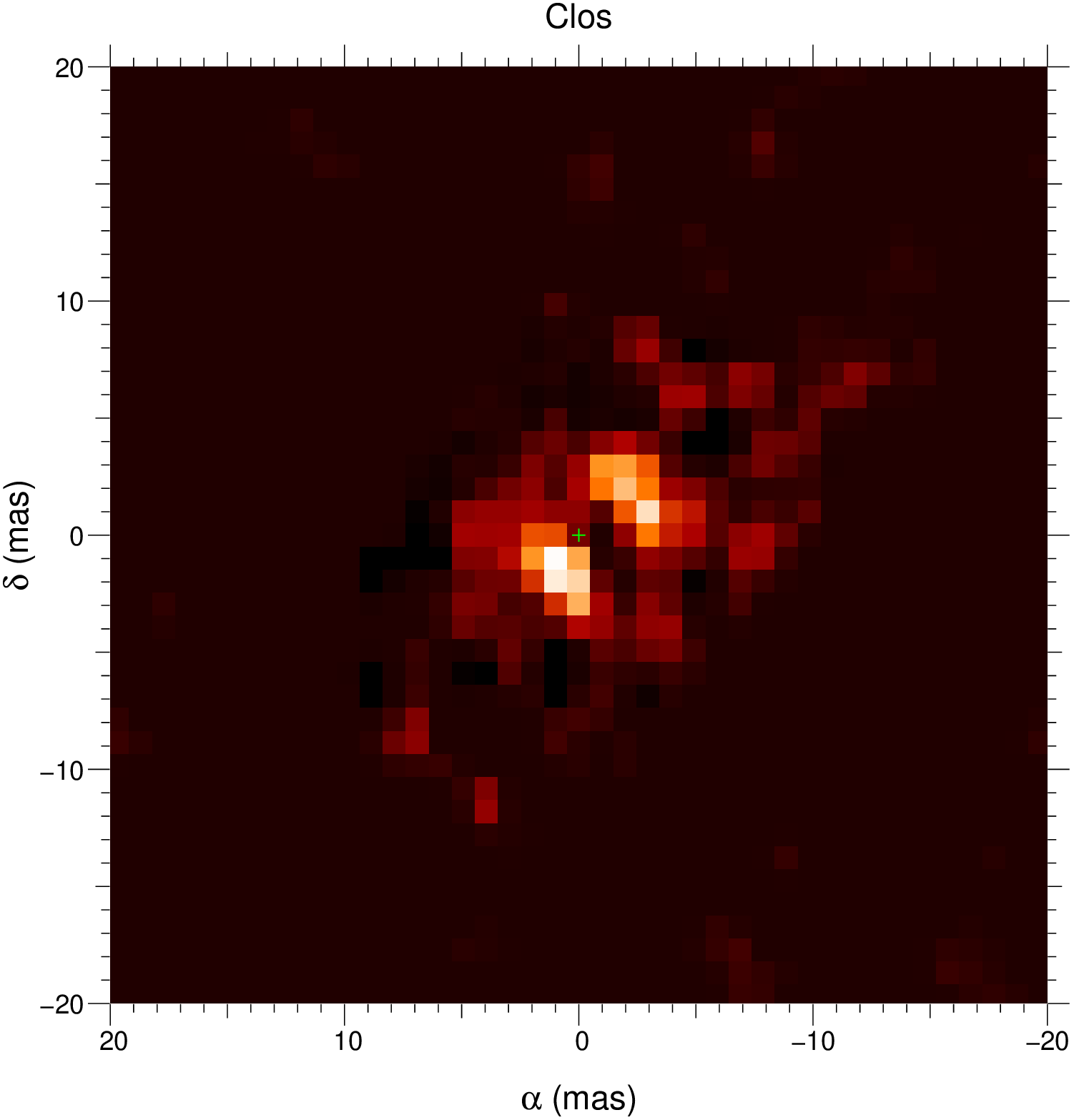}\\
\multicolumn{4}{c}{Image reconstruction (self-cal, step 1)}\\
\includegraphics[height=0.19\textwidth, angle=-90, origin=br]{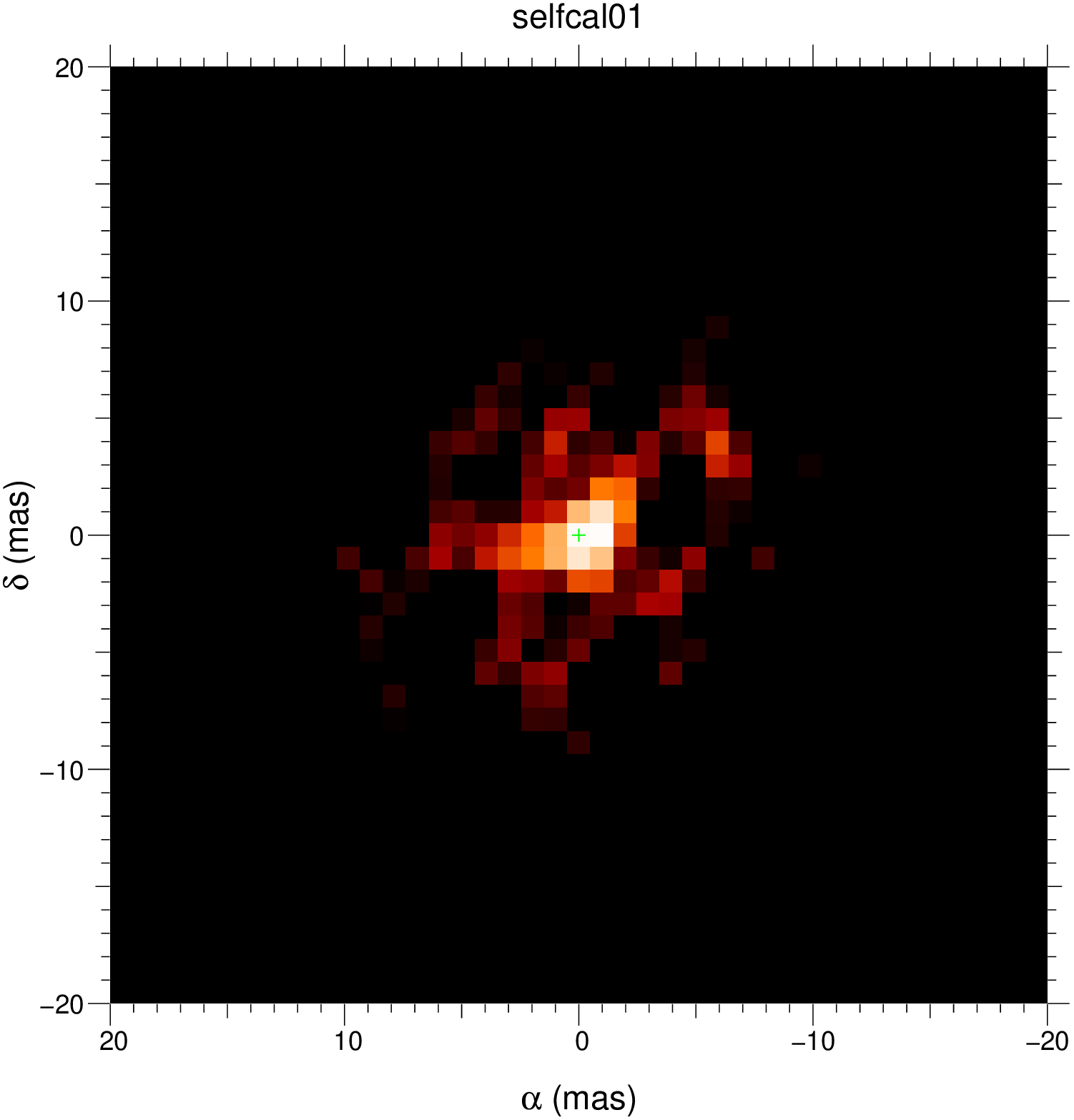}&
\includegraphics[height=0.19\textwidth, angle=-90, origin=br]{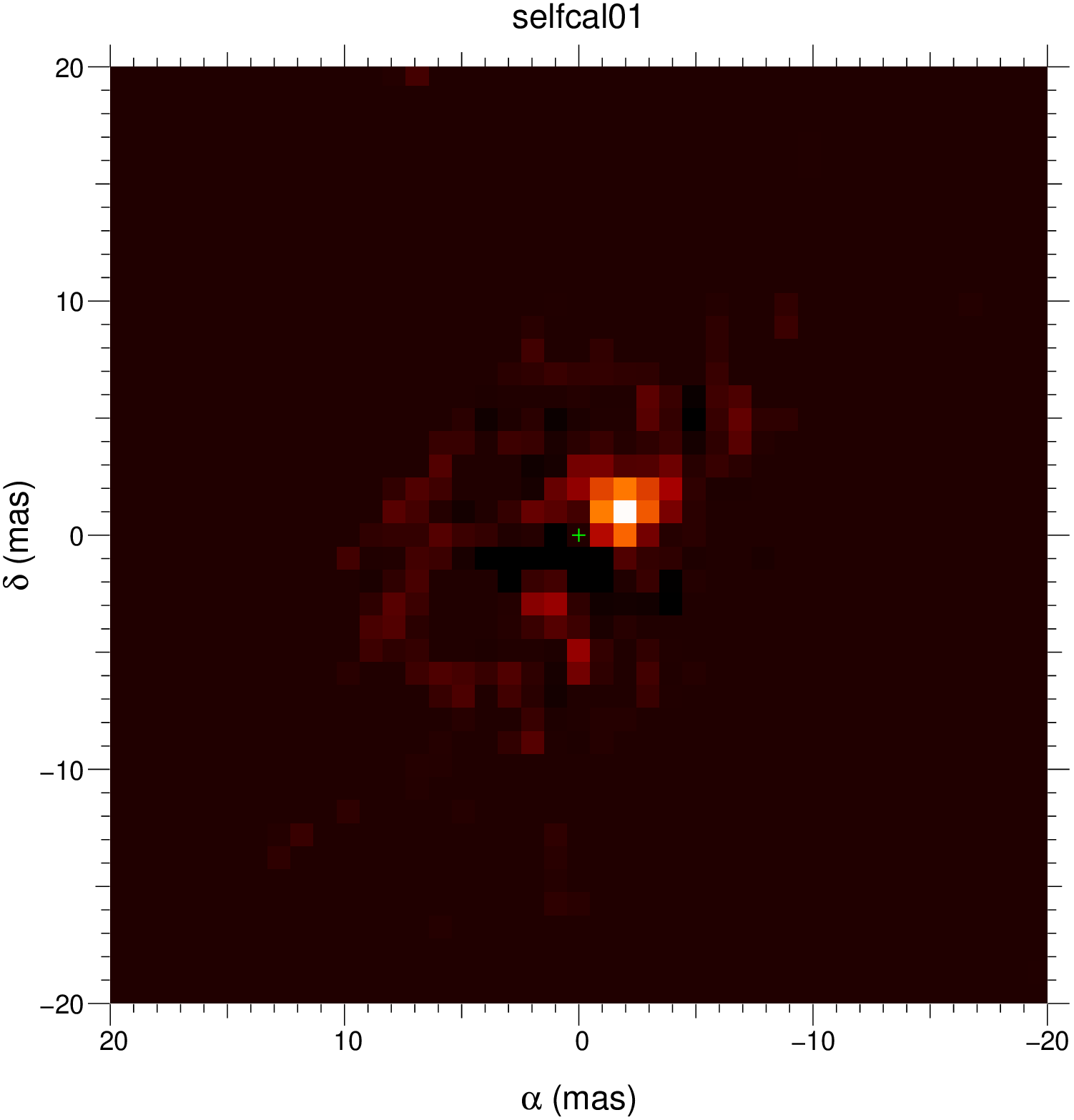}&
\includegraphics[height=0.19\textwidth, angle=-90,origin=br]{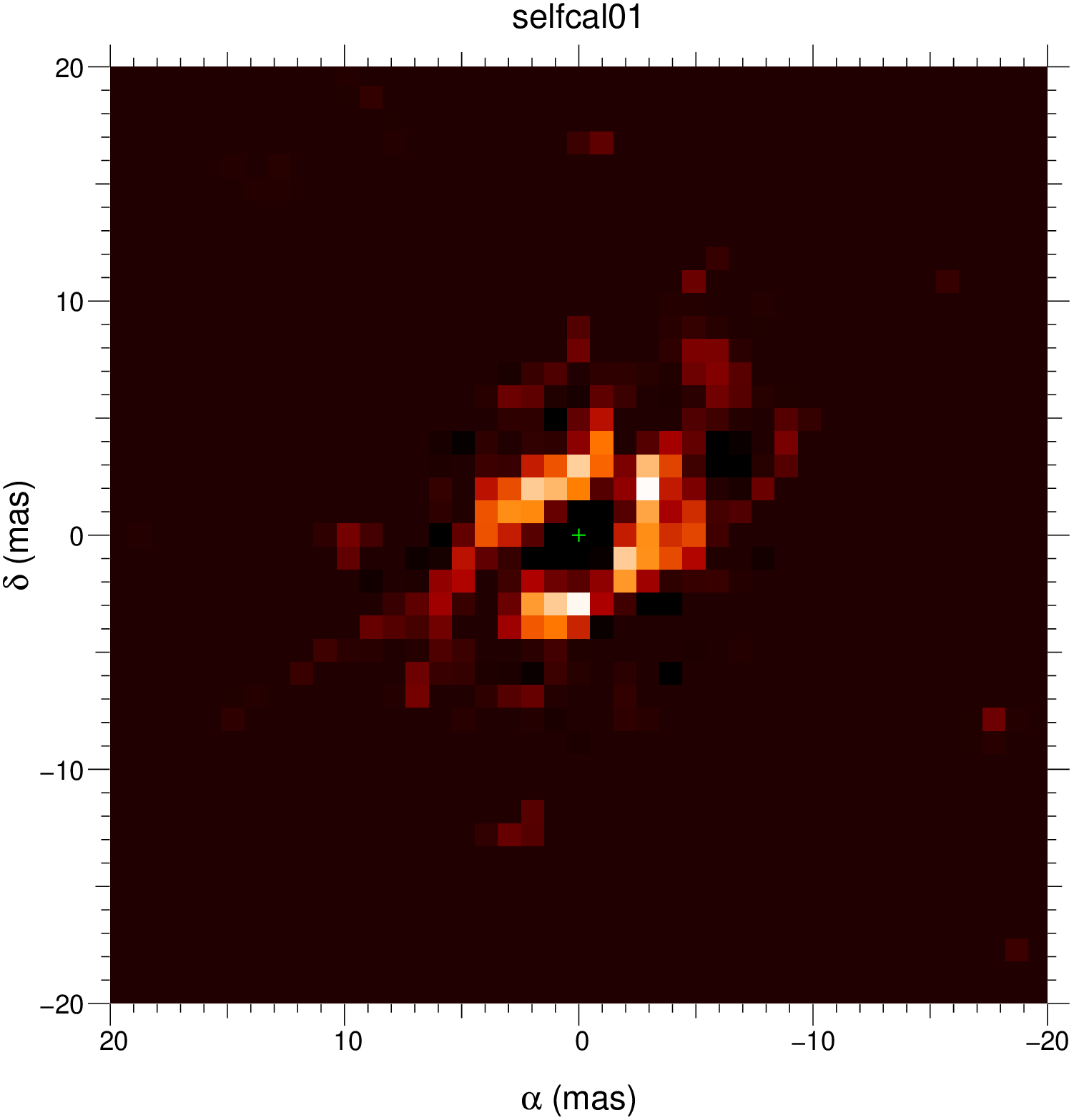}&
\includegraphics[height=0.19\textwidth, angle=-90,origin=br]{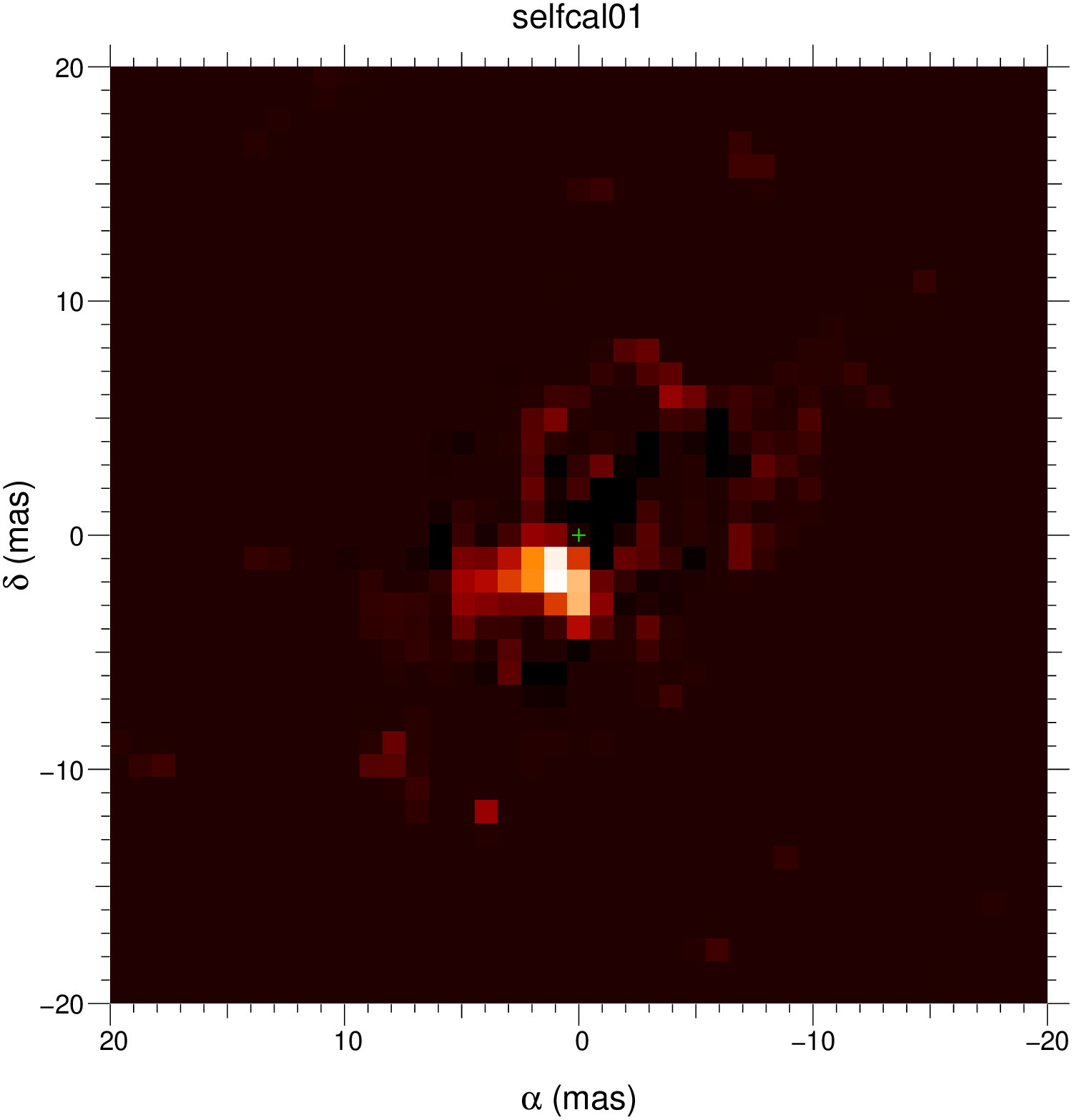}\\
\multicolumn{4}{c}{Image reconstruction (self-cal, after 5 steps)}\\
\includegraphics[height=0.19\textwidth, angle=-90, origin=br]{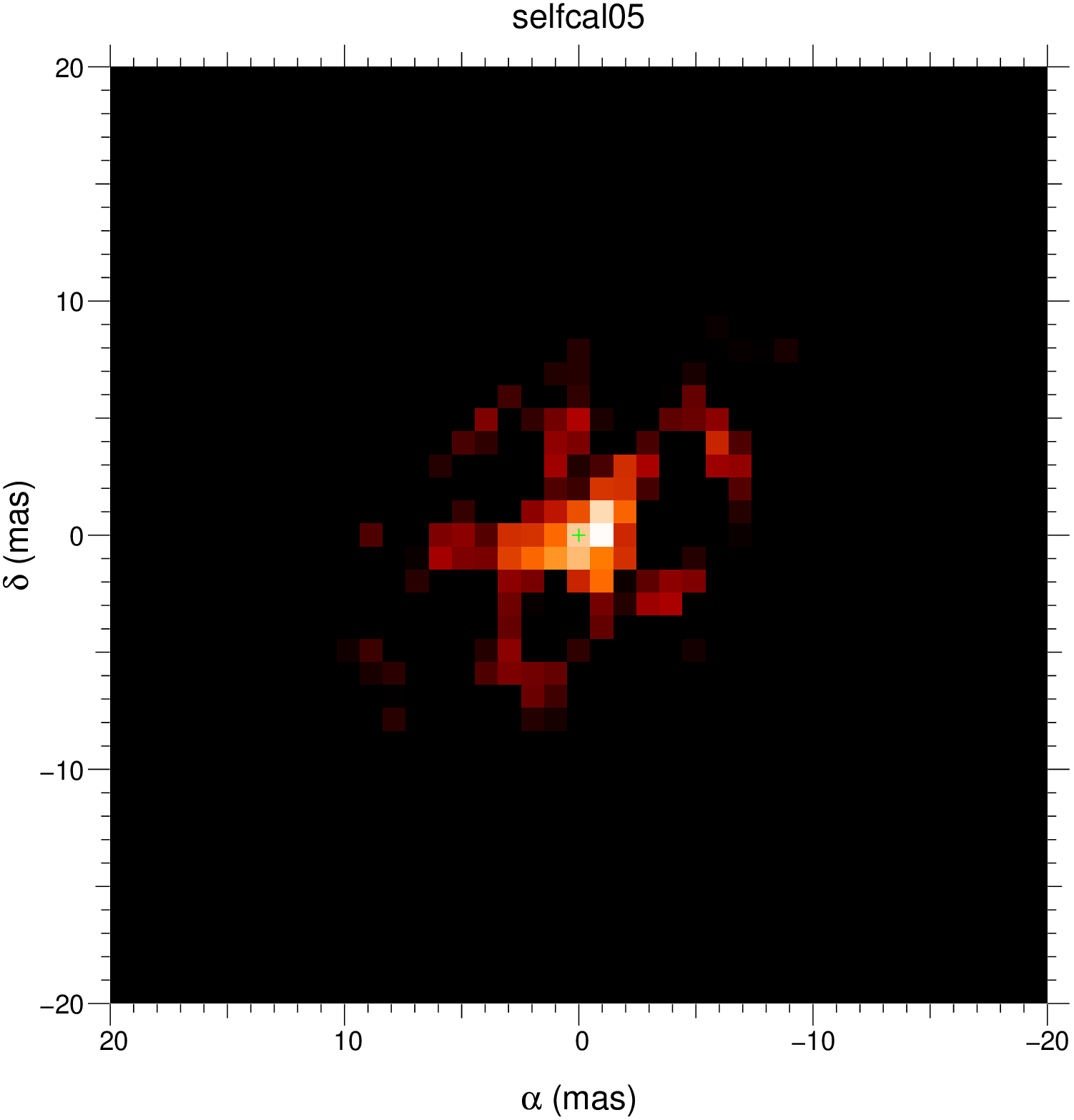}&
\includegraphics[height=0.19\textwidth, angle=-90, origin=br]{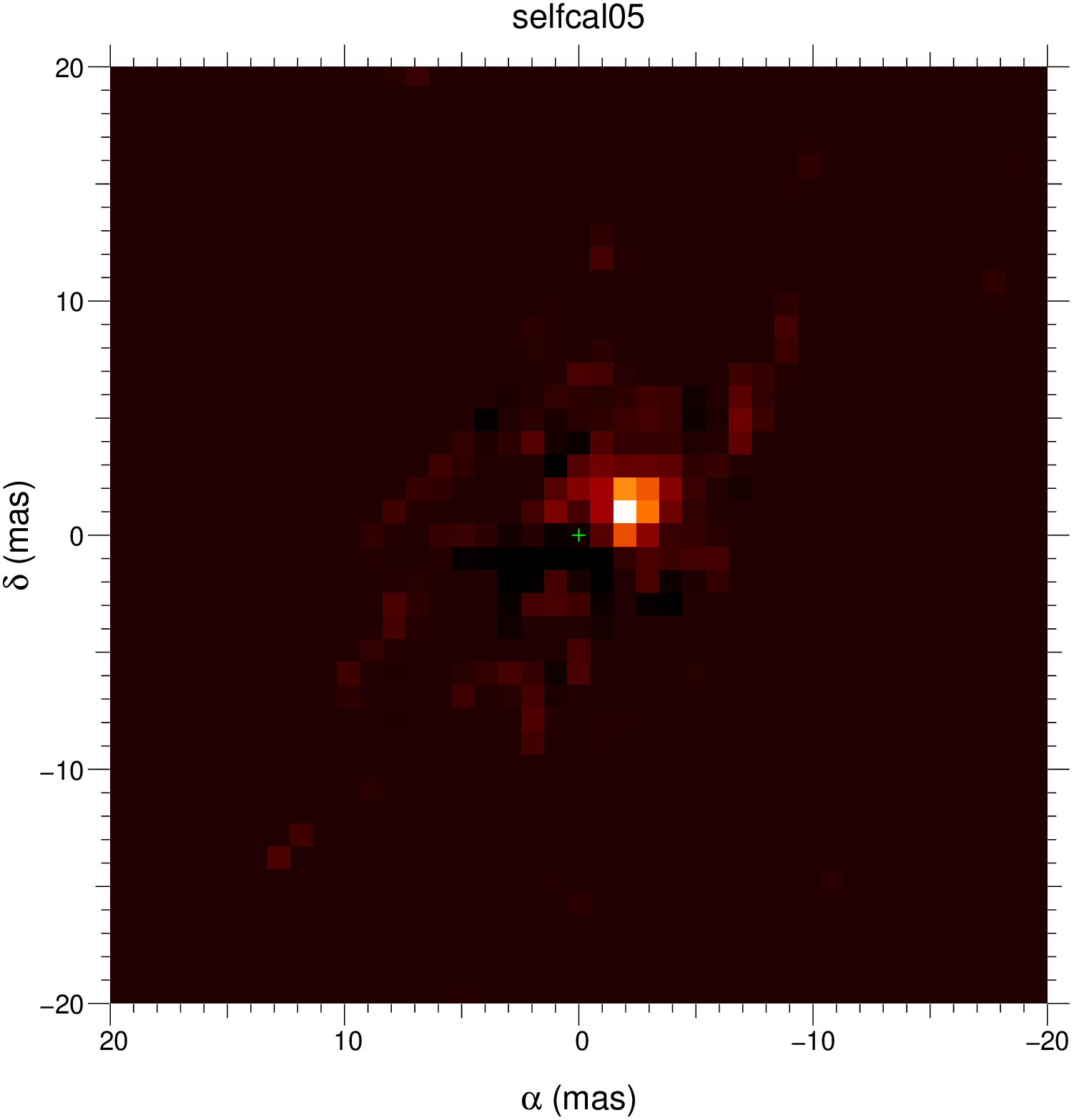}&
\includegraphics[height=0.19\textwidth, angle=-90,origin=br]{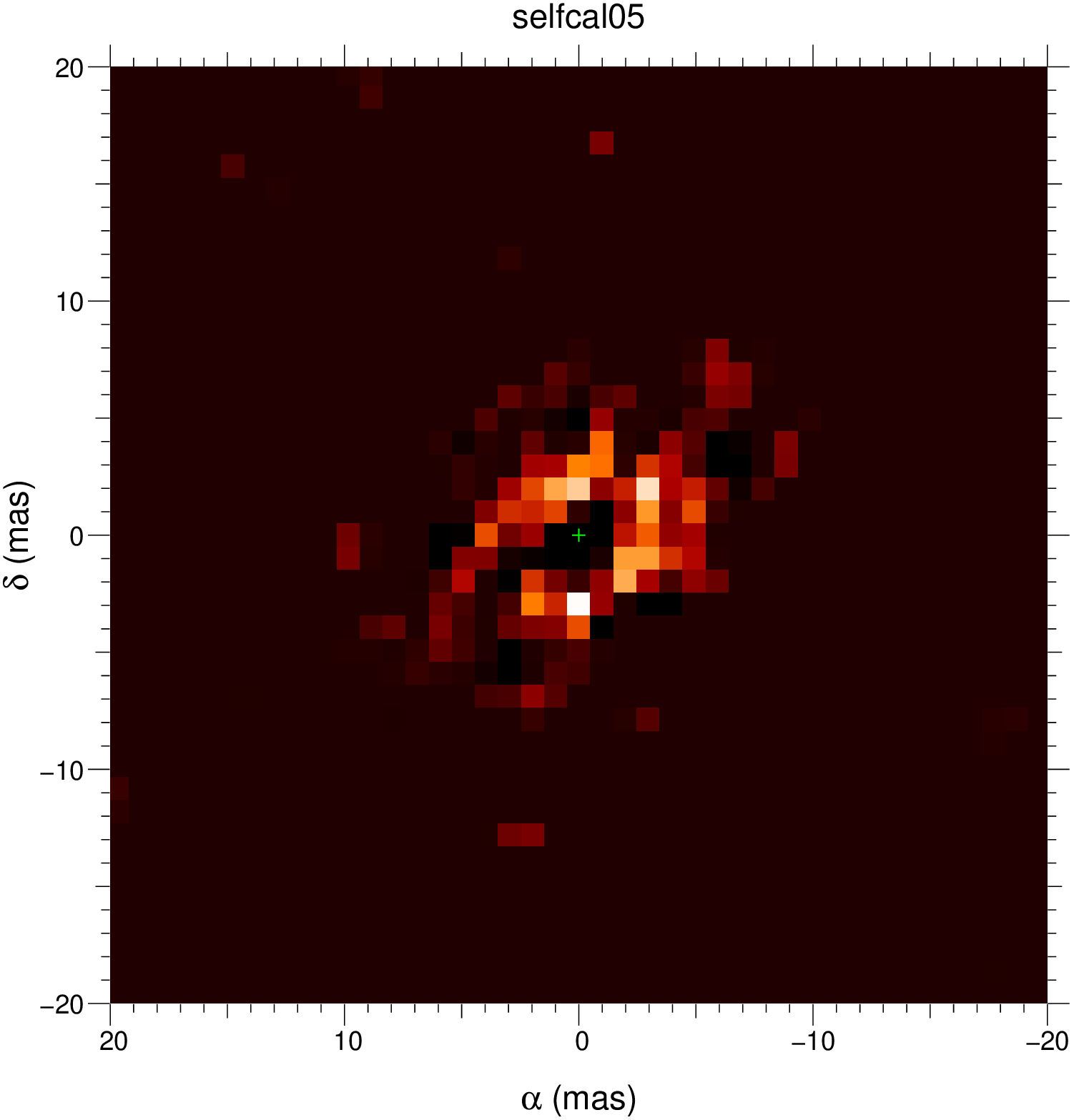}&
\includegraphics[height=0.19\textwidth, angle=-90,origin=br]{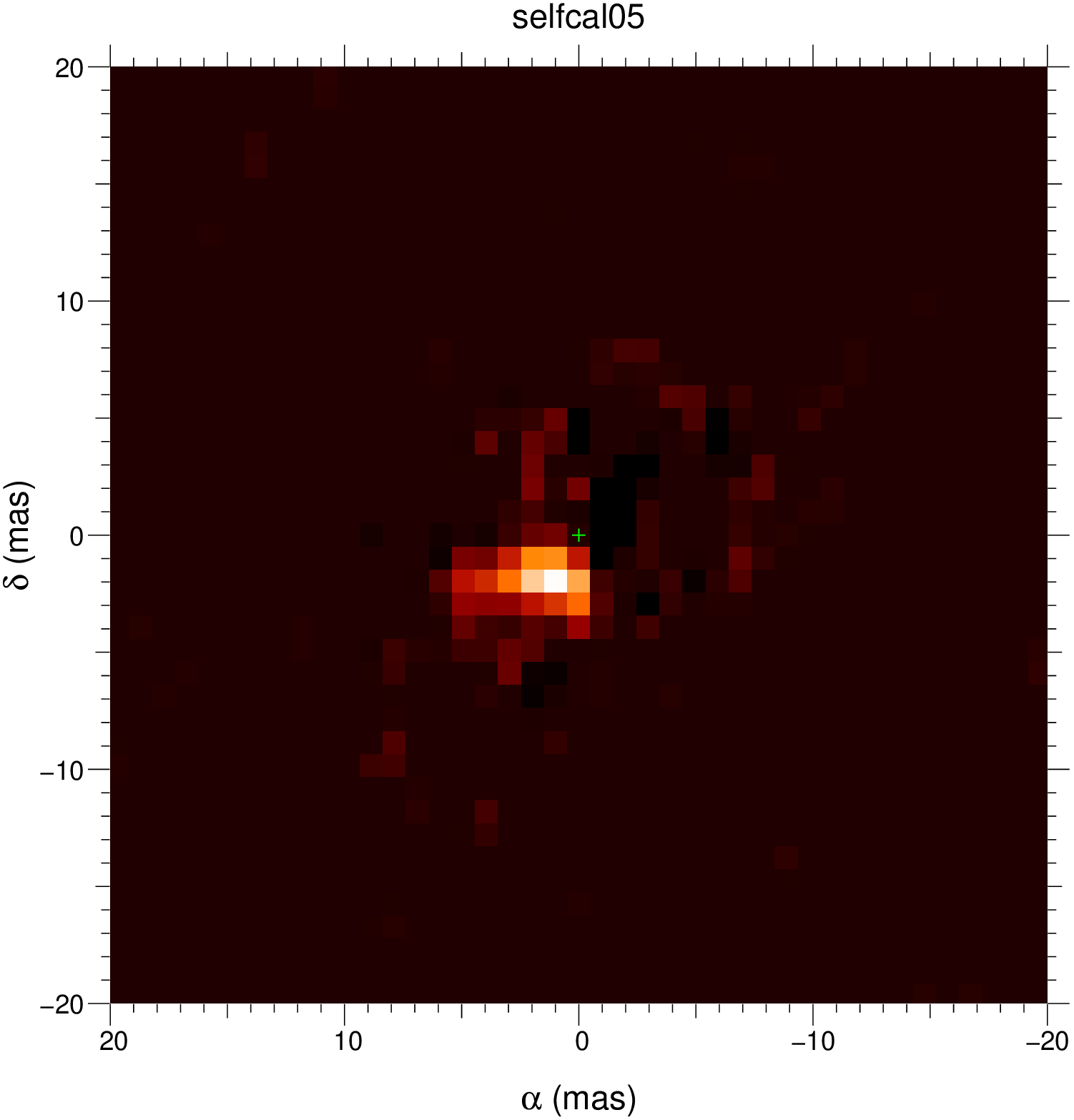}\\
\multicolumn{4}{c}{Image reconstruction (self-cal, after 10 step)}\\
\includegraphics[height=0.19\textwidth, angle=-90, origin=br]{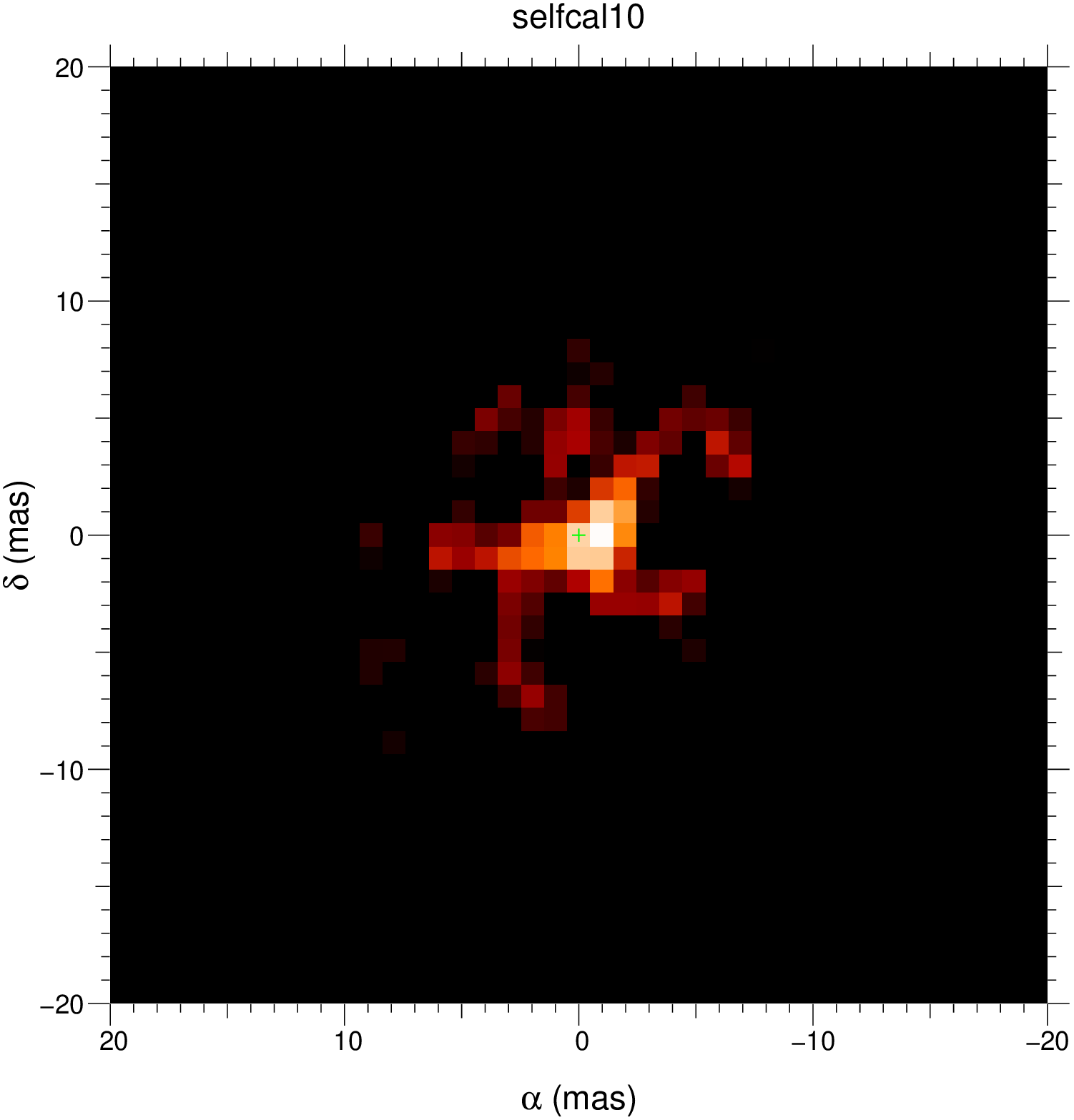}&
\includegraphics[height=0.19\textwidth, angle=-90, origin=br]{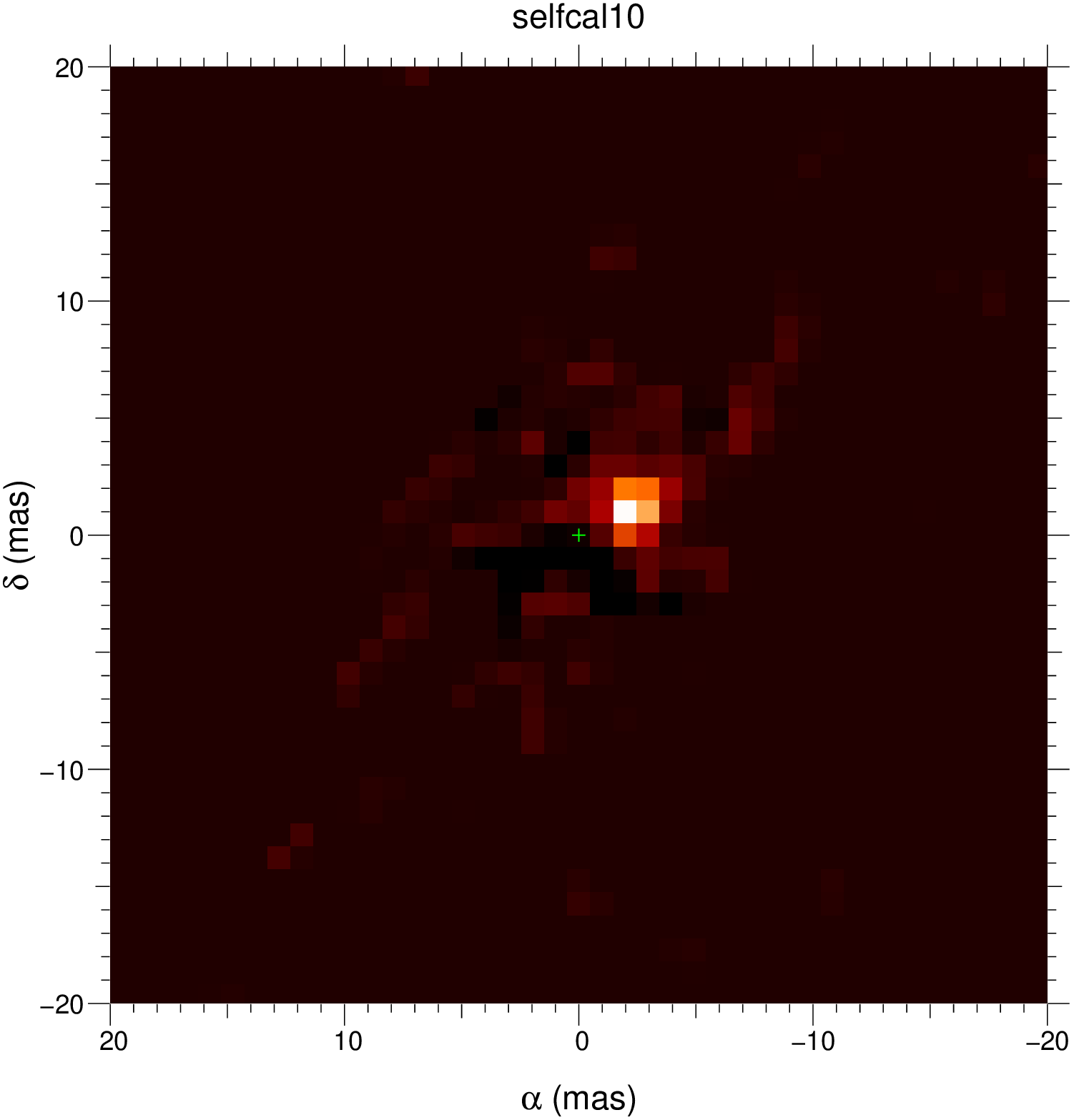}&
\includegraphics[height=0.19\textwidth, angle=-90,origin=br]{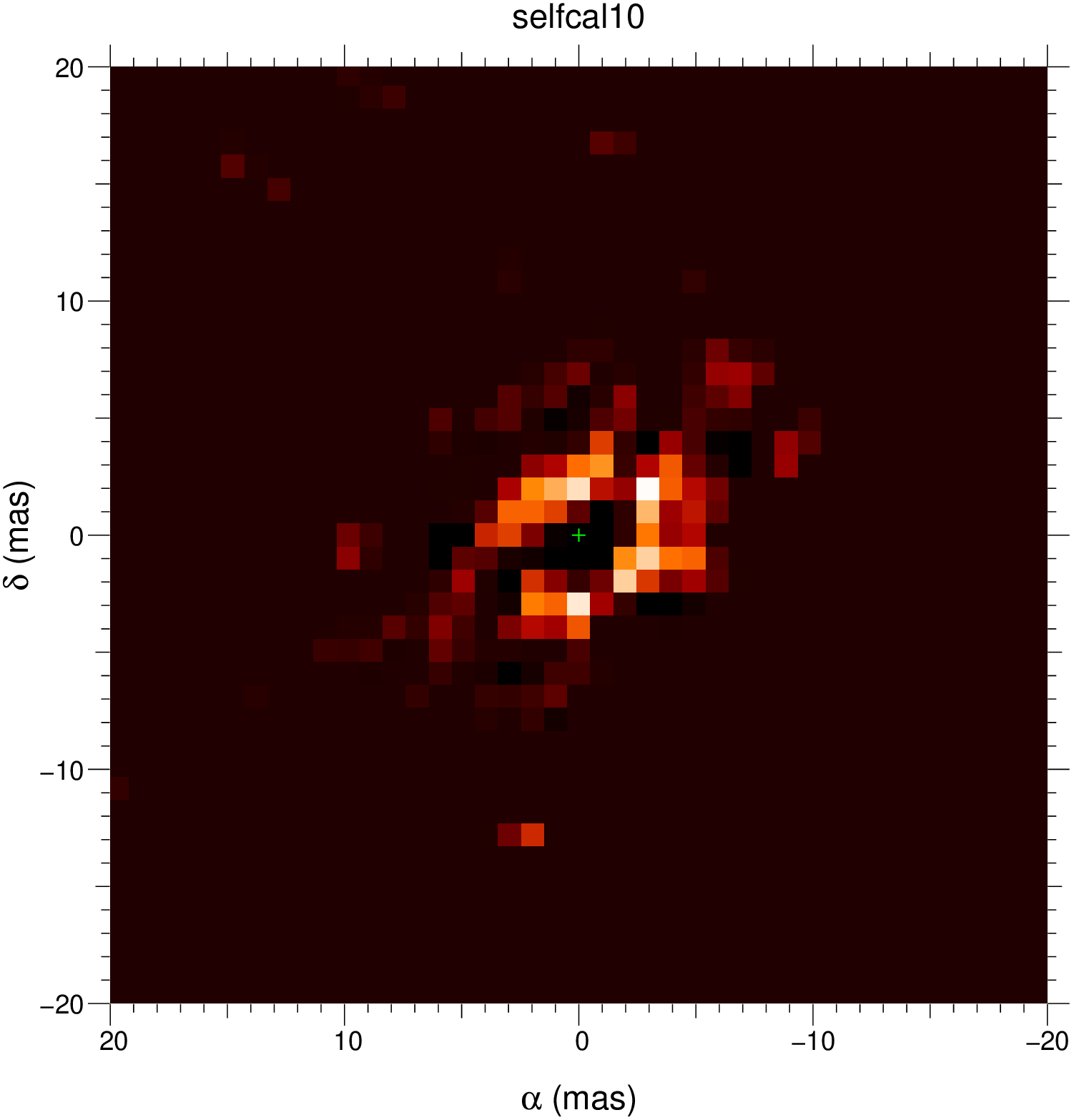}&
\includegraphics[height=0.19\textwidth, angle=-90,origin=br]{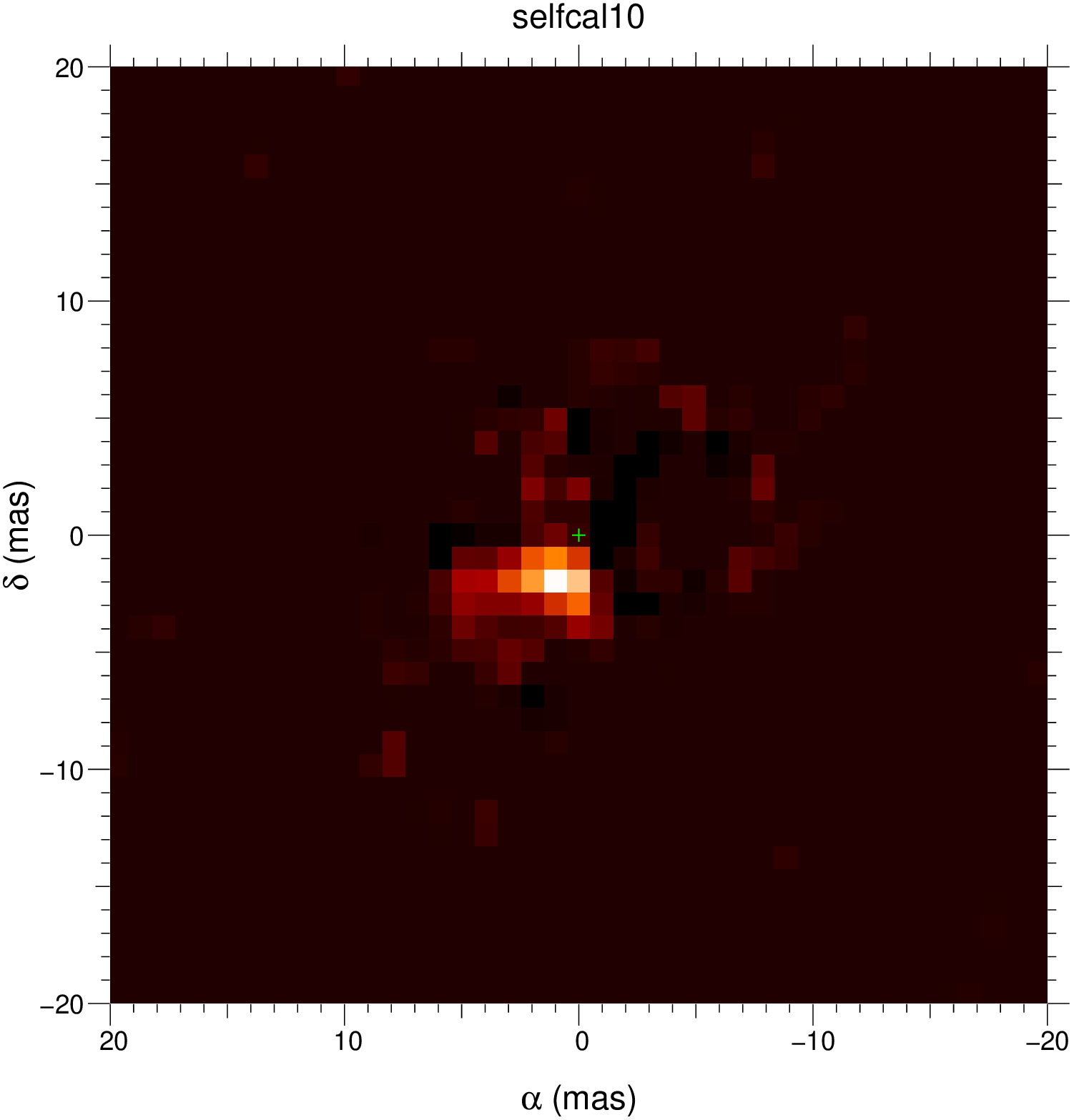}\\
\end{tabular}
\smallskip
\caption{Image reconstruction with Differential Phase Self-Calibration. The 4 images show from left to right the continuum image (with intensities as cubic root), and excess line emission in the blue wing, center and red wing of the emission line, respectively (intensities are plot linearly. The top row show the used model, the 2nd row the initial run with $V^2$ and closure phases, the 3\rd row the 1st step of self-cal, the 4\th row the 5\th step, then the 5\th row the 10\th step.}
\label{DPSC_Bestar}
\end{figure}

The simulation includes $V^2$, closure phases and the differential phases. A realistic noise model is applied to the data, and the image reconstruction is performed the same way as in Millour et al. 2011. Figure~\ref{DPSC_Bestar} displays the results of the image reconstruction at the different steps (initial reconstruction with $V^2$ and closure phase, self-cal step 1, 5 and 10). Similarly to what was found in Millour et al. 2011, we get a spectacular improvement of the image quality at the first step of self-cal. However, as seen also previously, we do not see an additional improvement in the following steps.

\section{Prospects} 

With the three presented recipes, we propose ways of improving the image reconstruction quality. While we made most of our tests using the MIRA software, BFMC and LFF can be applied to other image reconstruction software, like BSMEM or WISARD, as we experimented it. DPSC needs a software that can reconstruct images with complex visibilities. For the moment, only MIRA has been used to reconstruct images with this recipe, but in principle it can be applied to other software. Future improvements of this recipe will include a better propagation of errors and the inclusion of self-calibration of the visibility amplitudes in addition to differential phases.

 


\bibliography{millour2012}   
\bibliographystyle{spiebib}   

\end{document}